\documentclass[a4paper,fleqn,usenatbib]{mnras}
\usepackage{newtxtext,newtxmath}
\usepackage[T1]{fontenc}
\usepackage{ae,aecompl}
\usepackage{graphicx}	
\usepackage{amsmath}	
\usepackage{amssymb}	
\usepackage{pdflscape}

\newcommand{\lbolint}{\ifmmode L(IR;X-rays) \else $L${\it (IR; X-rays)}\fi}
\newcommand{\ledd}{\ifmmode L_{Edd}\else $L_{Edd}$\fi}
\newcommand{\lbol}{\ifmmode L_{BOL}\else $L_{BOL}$\fi}
\newcommand{\Mdot}{\ifmmode \dot{M} \else $\dot{M}$\fi}

 \newcommand{\nodata}{ ~$\cdots$~ }
 
\title[]{Infrared Colours and Spectral Energy Distributions of Hard X-ray Selected Obscured and Compton-thick AGN} 

\author[E. Kilerci Eser, T. Goto, T. Guver , A. Tuncer, O. H. Atas]{
Ece Kilerci Eser,$^{1}$\thanks{E-mail: ecekilerci@phys.nthu.edu.tw (EKE)}
T. Goto,$^{2}$
T. G{\"u}ver,$^{1}$
A. Tuncer,$^{1,3}$
O. H. Ata{\c{s}}$^{1,3}$
\\
$^{1}$Istanbul University, Science Faculty, Department of Astronomy and Space Sciences, Beyaz{\i}t, 34119, Istanbul, Turkey \\
$^{2}$Institute of Astronomy , National Tsing Hua University, No. 101, Section 2, Kuang-Fu Road, Hsinchu, 30013, Taiwan\\
$^{3}$Graduate School of Science and Engineering, Department of Astronomy and Space Sciences, Istanbul University, Istanbul, Turkey\\
}

\date{Accepted XXX. Received YYY; in original form ZZZ}

\pubyear{2020}

\begin{document}
\label{firstpage}
\pagerange{\pageref{firstpage}--\pageref{lastpage}}
\maketitle

\begin{abstract}
We investigate infrared colours and spectral energy distributions (SEDs) of 338 X-ray selected AGN from \textit{Swift}-BAT 105-month survey catalogue that have \textit{AKARI} detection,
in order to find a new selection criteria for Compton-thick AGN. 
By combining data from Galaxy Evolution Explore (\textit{GALEX}),  
Sloan Digital Sky Survey (SDSS) Data Release 14 (DR14), Two Micron All Sky Survey (2MASS),  Wide-field Infrared Survey Explorer (\textit{WISE}),  
\textit{AKARI} and \textit{Herschel} for the first time we perform ultraviolet (UV) to far-infrared (FIR) SED fitting 158 \textit{Swift}-BAT AGN by CIGALE 
and constrain the AGN model parameters of obscured and Compton-thick AGN. 
The comparison of average SEDs show while the mid-IR (MIR) SEDs are similar for the three AGN populations, optical/UV and FIR regions have differences. 
We measure the dust luminosity, the pure AGN luminosity and the total infrared (IR) luminosity. We examine the relationships between the measured infrared luminosities and the 
hard X-ray luminosity in the 14-195\,keV band. We show that the average covering factor of Compton-thick AGN is higher compared to the obscured and unobscured AGN. 
We present a new infrared selection for Compton-thick AGN based on MIR and FIR colours ([$9\micron - 22\micron] > 3.0$ and [$22\micron - 90\micron] < 2.7$) from  \textit{WISE}
and \textit{AKARI}. 
We find two known Compton-thick AGN that are not included in the  \textit{Swift}-BAT sample, and conclude that  MIR colours covering 9.7$\mu$m silicate absorption and MIR continuum
can be a promising new tool to identify Compton-thick AGN.

\end{abstract}




\section{INTRODUCTION}\label{introduction}

Active galactic nuclei (AGN) are the mass accreting supermassive black holes residing at the centres of nearly all massive galaxies. 
Recent studies show that  AGN that are obscured by gas and dust may make up a non-negligible fraction of the AGN population \citep[e.g.,][]{Hickox2007,Treister2010,Assef2013,Assef2015,Mateos2017}. 
Obscured AGN are important to understand the full growing black hole population and the influence of black holes on the host galaxy \citep[e.g.,][]{Hopkins2008}. 
These sources are important to understand the accretion history of super massive black holes via the cosmic X-ray background radiation \citep[e.g.,][]{Ueda2014}. 
It is still unknown how much obscured AGN are there in the Universe. 
The number of low luminosity or obscured AGN is critical for our understanding of galaxies \citep[e.g.,][]{Hickox2018}. Therefore, it is 
important to find new techniques to identify obscured AGN. 

The structure of an AGN mainly composed of a central super massive black hole, an accretion disc \citep{Shakura73} and a corona \citep[e.g.,][]{Czerny1987,Zheng1997,Telfer2002,Done2012,Jin2012a,Mehdipour2011,Mehdipour2015}. 
The accretion disc and the corona in its immediate vicinity produce  the primary optical-UV  and X-ray continuum emission \citep[e.g.,][]{Sanders1989,Marconi2004,Suganuma2006}. 
This central engine is embeded within the IR emitting dusty structure the so called torus \citep[e.g.,][]{Antonucci1993,Lawrence1991,Simpson2005,Honig2007,GarciaBurillo2016}.
AGN that can be identified by their blue continuum and broad and/or narrow emission lines in the optical are often referred as `Type 1' \citep[e.g.,][]{Antonucci1993}. 
Once the AGN signatures can not be detected in the optical spectrum, then its refereed as a `Type 2'  or obscured AGN  \citep[e.g.,][]{Antonucci1993}.
Obscured AGN may show narrow line emission or no emission at all depending on the degree of obscuration which is dictated by the geometrical structure of the dusty torus or 
the strength of the host galaxy emission \citep[e.g.,][]{Hickox2018}. Since they are mostly absorbed in the optical and soft X-rays, obscured AGN can mostly be identified in hard X-rays \citep[e.g.,][]{BrandtAlexander2015} and infrared (IR) \citep[e.g.,][]{Stern2005,Mateos2012,Assef2013,Hickox2017}.
In terms of the measured Hydrogen column density (N$_{\rm{H}}$) from the X-ray spectra AGN can be classified as: (i) unobscured ($\log$ N$_{\rm{H}} \leq 22.0$); (ii) obscured ($22.0 < \log$ N$_{\rm{H}} \leq 24.0$), and (iii) Compton-thick (CT) ($\log$ N$_{\rm{H}} \geq 24.0$). 

In the mid-IR AGN have characteristic emission, that is produced by the circumnuclear dust heated by the optical/UV/X-ray radiation from the central engine.
Therefore  it is a common practice to use IR emission to separate AGN from starburst/normal galaxies \citep[e.g.,][]{Stern2012,Mateos2013,Lansbury2014,Huang2017,Ichikawa2019}. 
The mid-IR dust emission of an AGN is in the form of power-law with  different slopes for Type 1 and Type 2 \citep{Alonso-Herrero2006,Donley2007,Donley2008}. 
 IR power-law selection has been used to select AGN with \textit{WISE} colours ([3.4\micron -4.6\micron] versus [4.6\micron - 12\micron]) for AGN at low redshift ($z\leq$0.5) \citep{Mateos2012,Mateos2013}. \textit{Spitzer} IRAC flux ratios ([8\micron -4.5\micron] versus [5.8\micron - 3.6\micron]) have been used as a reliable method to select AGN at higher redshift ($z\sim1, 2$) \citep{Lacy2004,Stern2005,Donley2008,Donley2012}.
Many AGN selection criteria have been shown and applied in other studies \citep{Jarrett2013,Stern2012,Assef2013}. Millions of AGN have been selected from AllWISE using the criteria of \citet{Mateos2012}.
The completeness of IR colour AGN selection is highly complete for luminous Type 1 AGN and moderately complete for Type 2 AGN \citep{Mateos2012,Lansbury2017}. It has been 
showed that \citep{Mateos2013,Rovilos2014} obscured AGN and CT AGN candidates meet the MIR-selection criteria of \citet{Stern2012,Mateos2013,Lansbury2014}.
These colours mainly identify obscured AGN; however this colour selection has been combined with N$_{\rm{H}}$ classification only in a few studies \citep{Mateos2013,Rovilos2014}.   
In this work we focus on selecting obscured AGN with a new IR colour criteria.

Since MIR and the X-rays radiation originate from the AGN  the two radiation is expected to be correlated. 
The relation between the mid-IR and X-ray emission has been an important tool to gain information about the AGN physics \citep[e.g.,][]{Krabbe2001,Lutz2004,Ramos-almeida2007, Horst2008, Fiore2009,Gandhi2009,Fiore2009,Levenson2009,Asmus2011,Mason2012,Sazonov2012, Matsuta2012, Ichikawa2012, Asmus2015,Mateos2015, Garcia-Bernete2016,Ichikawa2017,Chen2017,Ichikawa2019}. 
The mid-IR and X-ray relation initially established for the 2-10\,keV X-ray band, due to the technical limitations of the former X-ray telescopes. 
Once the hard X-ray telescopes like \textit{Swift} and \textit{INTEGRAL} become available the mid-IR and X-ray relation extended to the 14-195\,keV ultra-hard X-ray regime 
\citep[e.g.,][]{Mullaney2011,Matsuta2012,Ichikawa2012,Sazonov2012,Asmus2015,Ichikawa2017}.
\citet{Ichikawa2012} investigated the MIR and FIR properties of the AGN in the \textit{Swift}/BAT nine-month catalog \citep{Tueller2008} and find a good correlation between the MIR and X-ray luminosities. 
\citet{Ichikawa2017} and \citet{Ichikawa2019} studied the IR properties of the AGN in 70-month \textit{Swift}/BAT sample in a great detail. 
\citet{Ichikawa2019} analysed IR SEDs and estimated AGN contribution to the 12\micron, MIR, and total IR luminosities. 
They show a significant luminosity correlation between the MIR and 14-150\,keV bands. 
Although IR colours and properties of the hard X-ray selected AGN in the previous \textit{Swift}-BAT samples have been extensively studied, these literature studies \citet{Ichikawa2012,Ichikawa2017,Ichikawa2019} 
have not investigated a possible  Compton-thick AGN selection based on the  MIR and FIR colours which is the main goal of this study.

In this work, we combine the IR colours with the measured N$_{\rm{H}}$ of the X-ray selected AGN and 
present a new Compton-thick AGN selection criteria that allowed us to select  Compton-thick AGN candidates from the \textit{AKARI} all-sky survey catalogue. 
We also aim to constrain the AGN model parameters of obscured and CT AGN, within the limitations imposed
by the model assumptions in CIGALE, the available broadband photometry, and the fitting procedure. 
The structure of this paper is as follows. 
In sections \ref{S:sample} and  \ref{S:data} we present the sample selection and data, respectively. 
We describe sed analysis of obscured and CT AGN in our sample in section \ref{S:seds}.
We compare average SEDs of unobscured, obscured and CT AGN in \S \ref{S:avseds}. 
In section \ref{S:LxLir} we present the correlations between the hard X-ray luminosity and the pure AGN luminosity in the IR band and dust luminosity.
We show the N$_{\rm{H}}$ dependence of IR colours in section  \ref{S:colvsnh}. 
In section  \ref{S:selection} we present a new IR colour selection criteria for CT AGN.
We summarise our conclusions in \S \ref{S:conc}. 
We adopt a cosmology with $H_0=72$\,km\,s$^{-1}$\,Mpc$^{-1}$, $\Omega_\Lambda = 0.7$ and $\Omega_{\rm m}=0.3$ and use the base 10 logarithm.

\section{SAMPLE SELECTION} \label{S:sample}

The \textit{Swift} observatory \citep{Gehrels2004} scans the whole sky at 14$-$195\,keV with the wide-field Burst Alert Telescope \citep[BAT;][]{Barthelmy2005}. 
With its continuous all-sky survey, \textit{Swift}-BAT identified several hard-X-ray selected AGN in the local Universe \citep[e.g.,][]{Tueller2008,Baumgartner2013,Ricci2017,Oh2018}. 
The most recent data release of \textit{Swift}-BAT  \citep[105-month,][]{Oh2018} includes 947 hard X-ray selected non-beamed AGN. 
To investigate the mid-IR and far-IR colours of the hard X-ray selected obscured and CT AGN we select the 947 non-beamed AGN from \textit{Swift}-BAT 105-month survey catalogue \citep{Oh2018}. 
We cross-match these hard X-ray selected AGN based on their optical counter part coordinates with \textit{AKARI}/FIS all-sky survey bright source catalogue version 2\footnote[1]{http://www.ir.isas.jaxa.jp/AKARI/Archive/Catalogues/FISBSCv2/} \citep[][]{Yamamura2018} within a radius of 20 arcsec. 
We find \textit{AKARI} 90 $\mu$m detection for 378 AGN. 90 $\mu$m detection is near the peak of the far-IR dust emission
near 100 $\mu$m, and especially the 140 $\mu$m and 160 $\mu$m bands are crucial to constrain the dust emission peak and measure better IR luminosity. 
Among these AGN only 332 have detected in three \textit{AKARI} bands at 90 $\mu$m, 140 $\mu$m, and 160 $\mu$m. 
However, for our analysis when we require to use high quality \textit{AKARI} data with \textsc{fqual = 3} only 51 of these 332 AGN have high quality data in three  \textit{AKARI} bands. 
Therefore we use additional far-IR data when available as described in section \ref{S:data}.

Swift/BAT Spectroscopic Survey \citep[BASS;][]{Koss2017} DR1 release provides multiwavelength data of 836 local AGN from the \textit{Swift}-BAT 70-month catalogue \citep{Baumgartner2013}. The broad-band X-ray (between 0.3$-$150\,keV range) spectral analysis of the BASS AGN sample is completed by \citet{Ricci2017}. 
They combine  \textit{Swift}-XRT,  \textit{XMM}-\textit{Newton}, \textit{ASCA}, \textit{Chandra}, and \textit{Suzaku} observations for the soft X-ray (0.3$-$10\,keV) spectral analysis.
They fit the AGN continuum with a simple power-law model with Galactic absorption and use different models (see their table 2) when necessary to improve the fit. 
In order to measure the column density along the line-of-sight  \citet{Ricci2017} take into account both photoelectric and Compton scattering with \textsc{zphabs} and \textsc{cabs} models, respectively.
BASS includes 311 of the AGN in our \textit{AKARI} AGN sample, therefore for these sources we adopt the N$_{\rm{H}}$ values from BASS; except ESO244-IG030 and ESO317-G041 for which we adopt the 
revised N$_{\rm{H}}$ measurements from \citet{Marchesi2019}.
For 19 AGN in our \textit{AKARI} AGN sample we take the published N$_{\rm{H}}$ values from the literature \citep[][]{Maiolino1998,Fukazawa2001,Pappa2002,Zhang2006NH,Shu2007,Malizia2007,Malizia2008,Malizia2011,Malizia2012,Pian2010,deRosa2012,Matt2012,Balokovic2014,Gandhi2015,Gandhi2017,Koss2016,Giustini2017,She2017,Iwasawa2018,Marchesi2019}, based on the soft X-ray spectral analysis. 
We take X-ray data of 8 sources 
from public archives and perform the X-ray data analysis as described in Appendix \ref{S:Appendix}. 
In total 338 AGN in our sample have N$_{\rm{H}}$ measurements.

\begin{figure}
\begin{center}$
\begin{array}{c}
\includegraphics[scale=0.8]{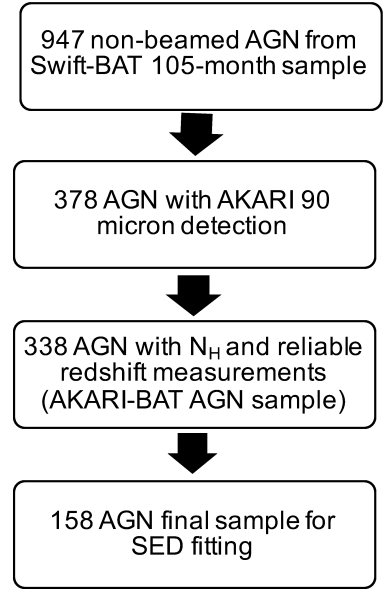} 
\end{array}$
\end{center}
\caption{ Flow-chart with all applied selection criteria for our \textit{AKARI}-BAT AGN sample. 
} 
\label{fig:fig1}
\end{figure}

\begin{figure}
\begin{center}$
\begin{array}{c}
\includegraphics[scale=0.8]{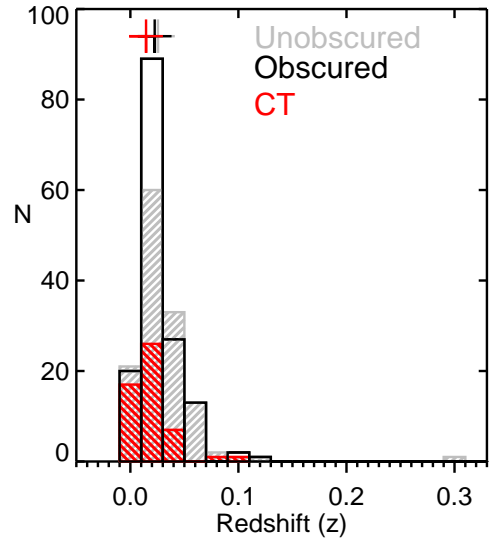} 
\end{array}$
\end{center}
\caption{ Redshift distribution of 338  hard X-ray selected \textit{AKARI} AGN sample. The median redshift of our sample is 0.023. 
The median redshifts (shown by the plus signs) of the unobscured, obscured and the CT AGN are  0.025, 0.023 and 0.014, respectively. 
For visual inspection the distribution is shown up to $z=0.33$, there are only two sources beyond this limit at $z=0.51$ and $z=0.60$.
} 
\label{fig:fig2}
\end{figure}

We focus our study on the hard X-ray selected \textit{AKARI}-BAT AGN sample that includes 338 sources with measured N$_{\rm{H}}$ values. 
A flow-chart describing the selection criteria of our final \textit{AKARI}-BAT AGN sample is displayed in Fig. \ref{fig:fig1}.
We list several source properties in Table \ref{tab:table1}.  
Based on the listed N$_{\rm{H}}$ values in Table \ref{tab:table1} our sample includes 133 unobscured, 153 obscured and 52 CT AGN. 
The redshift distribution of our sample of 338 AGN with reliable spectroscopic redshift measurements is shown in Figure \ref{fig:fig2}. 
The median redshift of our sample is 0.02. 

\begin{landscape}
\begin{table}
           \centering
           \small\addtolength{\tabcolsep}{-4pt}
           \caption{Hard X-ray selected \textit{AKARI} AGN sample. The full table is available in the electronic version of the article. 
           Columns:  (1) \textit{Swift} ID from \textit{Swift}-BAT 105-month survey catalogue. 
           (2) \textit{AKARI} ID from the \textit{AKARI}/FIS all-sky survey bright source catalogue version 2. 
           (3) and (4) Optical counterpart coordinates from \textit{Swift}-BAT 105-month survey catalogue. 
           (5) Redshift of the optical counterpart from \textit{Swift}-BAT 105-month survey catalogue.
            (6) Base 10 logarithm of the intrinsic Hydrogen column density in units of cm$^{-2}$. 
           (7) X-ray spectral analysis reference for the adopted $\log (N_{\rm{H}})$ measurement. 
           (8) Base 10 logarithm of the pure AGN luminosity in the IR band measured from the SED fitting by CIGALE.
           (9) The fractional AGN emission contribution to the total IR luminosity. 
           (10) The 14$-$195\,keV luminosity adopted from \textit{Swift}-BAT 105-month survey catalogue. 
           (11) \textit{WISE} 22 $\mu$m ($W4$) magnitude. 
           (12) $-$ (15) The \textit{AKARI} 9, 18, 65, 90 $\mu$m magnitudes from the \textit{AKARI}/FIS all-sky survey bright source catalogue version 2. and 
             \textit{AKARI}/IRC all-sky survey point source catalogue version 1. 
	}
           \label{tab:table1}
            \begin{tabular}{llrrrcccccccccc}
           \hline
            \textit{Swift} & \textit{AKARI}   & RA           & Dec         & $z$      & $\log (N_{\rm{H}})$ & X-ray & $\log (L(IR)_{AGN} $ &$frac_\mathrm{AGN}$ & $\log (L_{\rm{14 - 195}}$  &$F$(22) & $F$(9)  &$F$(18) & $F$(65) & $F$(90) \\
            source           & source           & (J2000.0) &(J2000.0) &            &                                & Ref.  &    / erg s$^{-})$          &                          & / erg s$^{-})$                 &  $\mu$m) & $\mu$m)&$\mu$m)&$\mu$m)&$\mu$m)\\
             name            & name             & (deg)        & (deg)      &            &                                 &         &                                   &                          &                                           &  (AB)              &(AB)&(AB) & (AB) & (AB)\\
            (1) & (2) & (3) & (4) & (5) & (6) & (7) & (8) & (9) & (10)  & (11) & (12) & (13) & (14) & (15)\\
           \hline
  J0002.5+0323  & 0002261+032111 & 0.6103 & 3.3519 & 0.0255 & 20.0$_{0.0}^{ 0.0} $& Bass  Survey & \nodata & \nodata & 43.20& 11.39  $\pm$ 0.03 & \nodata                  &  \nodata                  & 8.91  $\pm$ 0.27 & 9.08 $\pm$0.06\\
  J0006.2+2012 & 0006196+201211 & 1.5813 & 20.2029 & 0.0258 & 20.5 $_{0.2}^{ 0.0} $& Bass  Survey & 44.66& 0.6          & 43.38 & 10.19  $\pm$ 0.02 & 11.13  $\pm$ 0.03 & 10.53  $\pm$ 0.20 & 9.35  $\pm$0.34 & 10.17  $\pm$ 0.17\\
  J0038.4+2337 & 0038319+233652 & 9.6339 & 23.6133 & 0.0249 & 23.0$_{0.1}^{ 0.1}$ & Bass  Survey & \nodata&  \nodata & 43.26 & 12.10  $\pm$ 0.04 & \nodata                &   \nodata                 & 11.37  $\pm$ 1.18 & 10.04  $\pm$ 0.14\\
  J0042.9-2332 & 0042531-233224 & 10.7200 & -23.5410 & 0.0222 & 23.5$_{0.1}^{ 0.1} $& Bass Survey & 43.72 & 0.3           & 43.77 & 9.95  $\pm$ 0.02 & \nodata                   & 10.23  $\pm$ 0.11 & 8.08  $\pm$ 0.21 & 8.75 $\pm$ 0.07\\
  J0042.9+3016A& 0043022+301721 & 10.7578 & 30.2888 & 0.0520 & 22.3 $_{0.2}^{ 0.1}$& Bass  Survey & \nodata & \nodata & 43.60 & 11.32  $\pm$ 0.03 & \nodata               &   \nodata                 & 10.82  $\pm$ 0.95 & 10.00  $\pm$ 0.14\\
  J0051.6+2928& 0051348+292408 & 12.8959 & 29.4013 & 0.0360 & 20.0 $_{0.0}^{ 0.0}$& Bass  Survey & 44.24 & 0.1             & 43.49 & 11.41  $\pm$ 0.03 &\nodata             &  \nodata                  & 10.03  $\pm$ 0.61 & 8.74  $\pm$0.05\\
  J0105.5-4213& 0105260-421256 & 16.3617 & -42.2162 & 0.0302 & 24.2 $_{0.2}^{ 0.1} $& Bass  Survey &  \nodata &  \nodata & 43.48 & 10.94  $\pm$ 0.02 & \nodata               & 11.39  $\pm$ 0.26 & 9.42  $\pm$ 0.40 & 10.17  $\pm$ 0.15\\
		\hline
	\end{tabular}
\end{table}
\end{landscape}  

\section{DATA} \label{S:data}

To gain insight into the nature of our hard X-ray selected \textit{AKARI}-BAT AGN sample by investigating their spectral energy distributions (\S \ref{S:seds}) and colours (\S \ref{S:colvsnh} and \S \ref{S:selection}) we collect archival photometry from UV to far-IR. 

Galaxy Evolution Explorer \citep[\textit{GALEX}][]{Martin2005} performed a comprehensive sky survey at near-UV (NUV) and far-UV (FUV) bands at $\lambda_\mathrm{eff}=2267$\AA\  and $\lambda_\mathrm{eff}=1516$\AA, respectively. 
The revised catalog of \textit{GALEX} ultraviolet sources \citep{Bianchi2017} lists revised photometric measurements and eliminates any duplicate entries.
We cross match optical counterpart positions of AGN in our sample with the revised catalog of \textit{GALEX} ultraviolet sources within a  separation threshold
of 5 arcsec and find 205 UV counterparts. Since \citet{Bianchi2017} only include sources from All-Sky Imaging Survey (AIS) observations with both FUV and NUV 
detectors exposed, their catalogue do not include sources exposed only with a single FUV or NUV detector. To include such sources in our UV counterpart sample, 
the remaining 173 sources were cross matched (using a search radius of 5 arcsec) with GALEX source catalogue \citep{Martin2005} Data Release 6 (DR6)  from 
MAST Portal\footnote[2]{http://mast.stsci.edu}. After dropping matches with measurements with dichroic reflections ( \textsc{nuv\_artifact} = 4 ) and keeping only the single photometric 
measurement with the longest duration for  multiple observations of the same source, we find 56 more \textit{GALEX}  counterparts. In total we find 261 \textit{GALEX} counterparts.  
We correct the UV photometric measurements for Galactic foreground extinction using the listed $E_{B-V}$ values \citep[based on the extinction maps of][]{Schlegel1998} in the \textit{GALEX} catalogues. 

For the optical counterparts we extract the Galactic extinction corrected $u$-, $g$-, $r$-, $i$-, $z$-band magnitudes from the Sloan Digital Sky Survey \citep[SDSS;][]{York2000} Data 
Release 14 \citep[][]{Abolfathi2018} \textit{PhotoObjAll} catalog \footnote[3]{https://skyserver.sdss.org/dr14/}. We find only 146 sources in the SDSS DR 14 footprint. However 67 galaxies whose 
photometric flags include \textsc{clean} = 0 are not reliable and therefore not included in our analysis. 
We visually inspected images of the remaining galaxies and discarded any suspicious photometry of: (i)  very large (nearby) galaxies whose automated deblending is unreliable; 
(ii) too bright galaxies close to the r-band saturation magnitude limit at $r$ = 14.5 \citep{Strauss2002}; (iii) galaxies with contaminated by superposed stars.  
We use near-infrared Two Micron All Sky Survey \citep[2MASS;][]{Skrutskie2006} $J$-, $H$-, $K_{s}$-band magnitudes from the  Two Micron All Sky Survey Extended Source 
Catalog \citep[2MASS XSC;][]{Jarrett2000} and 2MASS Point Source Catalog \citep{Cutri2003}. 
For the 332 sources in the 2MASS XSC we use the `20mag/sq.' isophotal fiducial elliptical aperture magnitudes. 
For the 17 sources in the 2MASS Point Source Catalog  we use default magnitudes. 
We check photometry artifact contamination and/or confusion flag (cc\_flag) and use only realiable measurements.
2MASS magnitudes were corrected for Galactic extinction based on the \citet{Schlegel1998} maps and the extinction law of \citet{Cardelli1989}.

In order to include \textit{AKARI} and \textit{WISE}  mid-IR photometry at 3.4, 4.6, 9, 12, 18 and 22 $\mu$m, we cross-match \textit{AKARI}/FIS coordinates of our hard X-ray selected \textit{AKARI} AGN sample with  the \textit{AKARI}/IRC all-sky survey point source catalogue version 1\footnote[4]{http://www.ir.isas.jaxa.jp/AKARI/Observation/PSC/Public/RN/AKARI-IRC\_PSC\_V1\_RN.pdf} and the AllWISE source catalogue\footnote[5]{http://wise2.ipac.caltech.edu/docs/release/allwise/} 
\citep{Cutri2013} using a search radius of 20 arcsec. 

\textit{Herschel} Space Observatory \citep[\textit{Herschel};][]{Pilbratt2010} mapped about a small fraction of the sky in far-infrared and sub-millimeter  bands centred at 70, 100, 160, 250, 350 and 500 $\mu$m
with the Photodetector Array Camera and Spectrometer  \citep[PACS;][]{Poglitsch2010} and Spectral and Photometric Receiver \citep[SPIRE;][]{Griffin2010} instruments. 
\citet{Melendez2014} and \citet{Shimizu2016} present \textit{Herschel} PACS and SPIRE photometric measurements for 313 AGN from the 58 month 
\textit{Swift}/BAT catalog \citep{Baumgartner2013}, respectively. 
By matching the optical counterpart names with \textit{Herschel}-BAT AGN sample of  \citet{Melendez2014} and \citet{Shimizu2016} 
we find \textit{Herschel} PACS photometry for 155 and SPIRE photometry for 187 AGN in our sample. 

\begin{figure*}
\begin{center}$
\begin{array}{cc}
\includegraphics[scale=0.7]{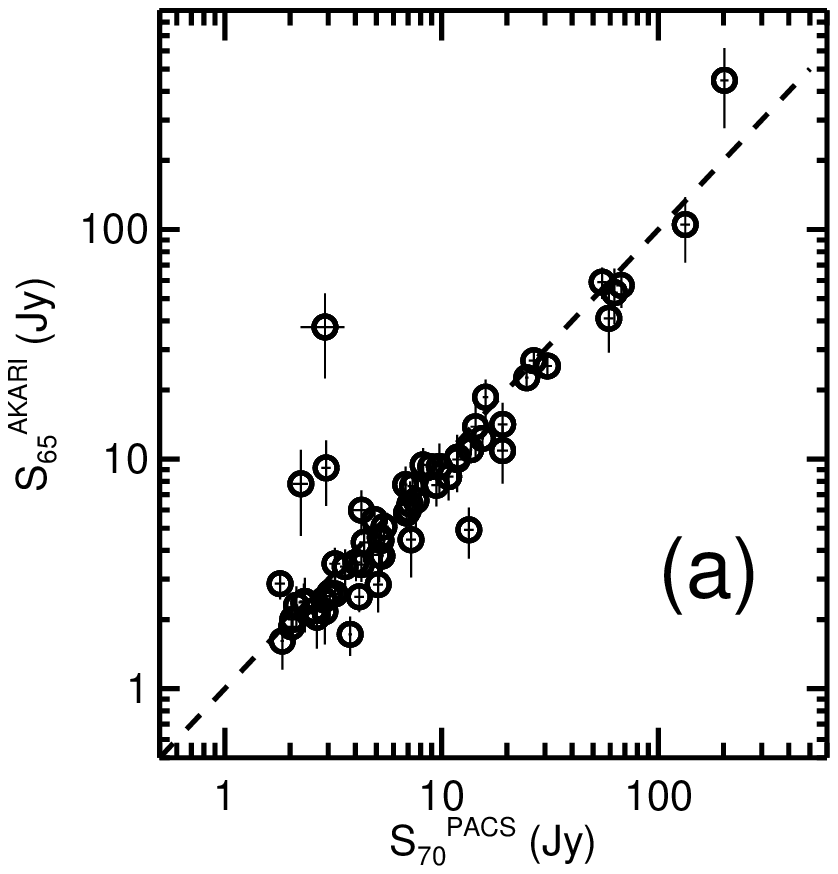} &
\includegraphics[scale=0.7]{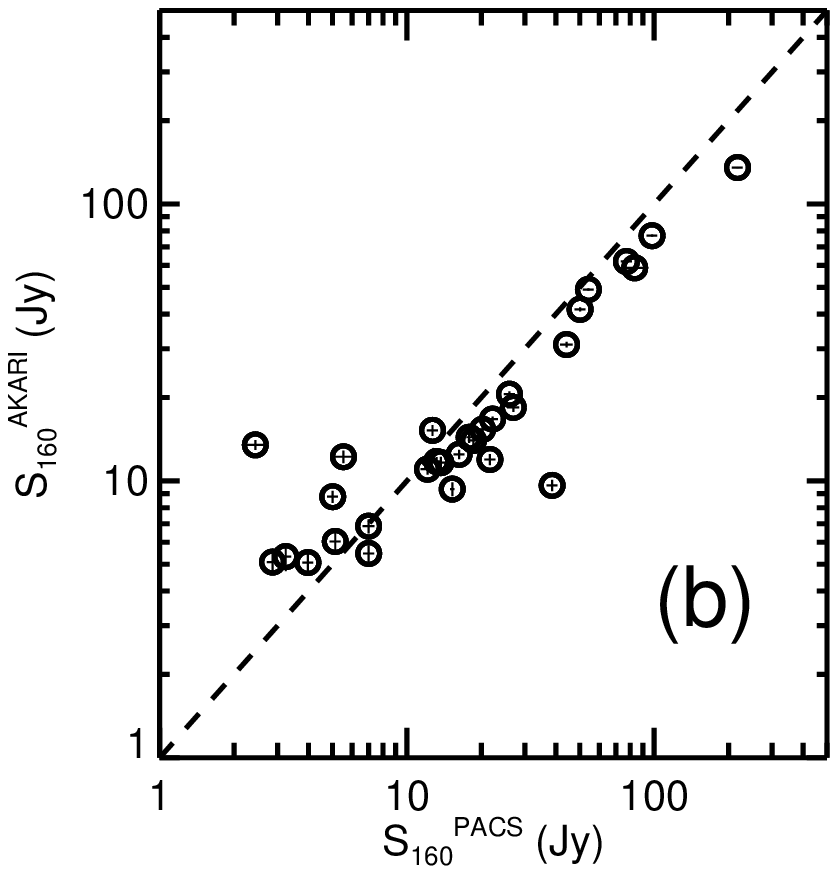} \\
\end{array}$
\end{center}
\caption{Comparison between the PACS and \textit{AKARI} fluxes at 70 $\mu$m$-$65 $\mu$m (a) and 160 $\mu$m$-$160 $\mu$m (b). The dashed lines represent $y=x$. 
} 
\label{fig:fig3}
\end{figure*}

When photometric fluxes from  \textit{AKARI}, \textit{WISE} and \textit{Herschel} are combined for the SED analysis (\S \ref{S:seds}) the spatial resolution differences among each band 
may cause nuclear and host galaxy IR emission measurement discrepancies \citep[e.g.][]{Clements2019}. 
The beam size for the \textit{AKARI}/FIS is \citep[beam FWHM is $\sim 30 - 50$ arcsec,][]{Doi2015}  larger than that 
of \textit{Herschel}/PACS \citep[beam FWHM is $\sim 6 - 11$ arcsec,][]{Poglitsch2010}. 
As shown by \citet{Clements2019} when \textit{AKARI} FIR fluxes are compared to \textit{IRAS} fluxes, the $\sim 10$ times better angular resolution of \textit{AKARI} causes missing FIR emission for extended sources. 
\citet{Clements2019} derive beam corrections  that should be applied to \textit{AKARI}/FIS all-sky survey bright source catalogue version 2 fluxes at 65$\mu$m and 90\,$\mu$m. These corrections depend 
on the extendedness parameter measured as the $J$ band magnitude difference between the 2MASS point source catalogue and 2MASSX extended source catalogue. 
We apply this beam correction \citep[see equation 2 of][]{Clements2019} to the 65$\mu$m and 90$\mu$m \textit{AKARI}/FIS fluxes in our sample. 
The applied beam correction results an average of 4\% and 20\% additional flux at  65$\mu$m and 90$\mu$m, respectively. 
\citet{Clements2019} also find that no beam corrections are needed for the moderately extended sources at 140$\mu$m and 160$\mu$m bands.  

As reported by \citet{Mushotzky2014} bulk of BAT sources are point like with a compact host galaxy with \textit{Herschel} spatial resolution. 
The comparison of  aperture correction applied (as described in the above) \textit{AKARI}/FIS  65$\mu$m ($S_{65}^{AKARI}$) and \textit{Herschel}/PACS 70$\mu$m ($S_{70}^{PACS}$) fluxes for 
61 sources in our sample show a a good agreement (Fig. \ref{fig:fig3}), the median difference between the two fluxes is 18 per cent with respect to $S_{65}^{AKARI}$. 
A typical difference of 18 per cent between the two fluxes is added in the quadrature to the flux uncertainties of the $S_{65}^{AKARI}$ fluxes. 
Since both $S_{70}^{PACS}$ and  $S_{65}^{AKARI}$ fluxes agree, we use both as independent photometric measurements in our SED analysis. 
30 sources in our sample have both high quality \textit{AKARI} and PACs fluxes at 160 $\mu$m band. 
As shown in  Fig. \ref{fig:fig3}, the median percentage difference between \textit{AKARI} and PACS 160 $\mu$m fluxes is 32\% with respect to the \textit{AKARI} fluxes. 
As reported by \citet{Mushotzky2014} the FIR radiation of the most BAT sources are point-like at the spatial resolution of \textit{Herschel}. Additionally, \citet{Melendez2014} and \citet{Shimizu2016} present PACs measurements  for an appropriate aperture for each source, therefore we prefer to use PACs 160 $\mu$m fluxes for these 30 AGN. 
If \textit{Herschel}/PACs 160 $\mu$m band photometry is not available we use AKARI/FIS 160 $\mu$m flux measurements for the rest of our sample. 
Since both \textit{AKARI}/FIS 140\,$\mu$m and 160\,$\mu$m bands are similar, to be conservative  we add a 32\% flux uncertainty in the quadrature to the photometric uncertainties of the 140\,$\mu$m and 160\,$\mu$m fluxes.

The 9 arcsec beam size of \textit{AKARI}/IRC is \citep{Ishihara2010} similar to the beam size of \textit{WISE} 
that is between 6 arcsec and 12 arcsec \citep{Wright2010}. 
We only use \textit{WISE} measurements with zero $cc\_flags$ values ($cc\_flags='0000'$) to avoid contaminated measurements (i.e., by diffraction spikes, bright sources). 
For extended \textit{WISE} sources (with $ext\_flg > 0$) we use elliptical aperture magnitudes (`$wngmag$', $n$ is the band number). 
For \textit{AKARI}/IRC photometry we only include measurements with high quality ($FQUAL(9\,\mu$m$) = 3$, $FQUAL(18\,\mu$m$) = 3$). 
\textit{AKARI}/IRC 9\,$\mu$m, 18\,$\mu$m and  \textit{WISE} 12\,$\mu$m, 22\,$\mu$m  bands have close but slightly 
different central wavelengths.
In our sample 112 sources have both \textit{AKARI}/IRC 9\,$\mu$m and \textit{WISE} 12\,$\mu$m fluxes. 
Additionally, 139 sources have both \textit{AKARI}/IRC 18\,$\mu$m and \textit{WISE} 22\,$\mu$m measurements. 
 For these sources, the comparison of \textit{AKARI}/IRC and \textit{WISE} fluxes agree (the mean percentage difference $\sim$20\%) without any systematic differences. 
 Therefore, we do not apply any aperture correction at these bands and use each photometric measurement separately in our SED analysis. 
However, to be conservative we add the typical flux difference of $\sim$20\% in the quadrature to the photometric uncertainties of the 9\,$\mu$m, 12\,$\mu$m,  18\,$\mu$m and 22\,$\mu$m fluxes.

\section{Infrared Spectral Energy Distributions of Unobscured Obscured and Compton-thick AGN} \label{S:seds}  

We perform SED fitting based on the collected photometric data by using the Code for Investigating Galaxy Emission \citep[CIGALE;][]{Noll2009,Serra2011,Boquien2019} 
version 2018.0. CIGALE\footnote[6]{http://cigale.lam.fr/.}  is a modern galaxy SED modelling code that applies energy conservation principle between the 
near-infrared/optical/ultraviolet emission that is absorbed by dust and the re-emitted mid-infrared and far-infrared emission. CIGALE models host galaxy emission 
and AGN component separately. CIGALE combines several built models based on the given input parameters for stellar, dust and AGN components. 
As a result it generates the probability distribution function of the model parameters. The mean value of the probability distribution function is the output value of a parameter. 
The standard deviation measured from the probability distribution function is the associated error of the parameter. 

Stellar component models include the stellar population, Initial Mass function (IMF) and star formation history. 
Here we adopt the stellar population models of  \citet{Maraston2005} with Salpeter IMF \citep{Salpeter1955} and a double exponential star formation history. 
For the dust component we use the dust models of \citet{Draine2014} and for the dust attenuation we use modified \citet{Charlot&Fall2000} attenuation law. 
We model the AGN component with the models from \citet{Fritz2006}. 
These AGN models are composed of the isotropic central source emission  in the form of power laws between 0.001$-20\mu$m  and the dust emission of the toroidal obscurer. 
In these models the central AGN emission can be partly absorbed by the dust and re-emitted at 1-1000$\mu$m  or scattered by the dust. 
In Table \ref{tab:table2} we list the adopted parameters used for the SED fitting. 
Table \ref{tab:table2} includes 17 parameters, but we note that some of the parameters and data are correlated and interconnected. As a result, the number of degrees of freedom is not 17. CIGALE compares one dataset (with N observed data points) to every single model. So the Bayesian method used in CIGALE do not estimate each parameter separately but measure the likelihood of each model (built from a set of parameters) to match the dataset. And from these likelihoods, the probability are evaluated and the probability distribution functions are built.

In CIGALE the quality of the parameter estimation process depends on the provided data points at different regions over the SED. 
For example, CIGALE uses UV-optical and  near-infrared data to model stellar component. 
Additionally mid-infrared and far-infrared data are needed to model the dust component. 
To estimate reliable luminosities from the SED fitting, we require to have at least one detection at each  UV-optical, near-infrared, mid-infrared and far-infrared regions. 
For the far-infrared part, to accurately constrain the dust SED we require to have at least one detection at shorter wavelengths  than the peak of the dust SED near 100$\mu$m and 
at least one detection at longer wavelengths.
Table \ref{tab:table3} list the broad-band filters included in our SEDs, the detection rate of our sample of 338 AGN at each band and the number of sources detected at different wavelength regions of the SED. 
The data requirement for the SEDs ensures all of the SEDs to have at least 5 data points. 
As a result of the applied wavelength coverage criteria we perform SED fitting analysis for 158 AGN in our sample. Among those 68 are unobscured,  65 are obscured and 25 are CT AGN. 

\citet{Fritz2006} model the AGN emission by considering the radiative transfer model of three main emission components.  These are: (i) the central illuminating source; 
(ii) the scattered emission by dust in the torus; (iii) the thermal dust emission from the torus. These models depend on seven parameters. The size of the torus, $r$ is defined by 
 ratio of the maximum outer radius to minimum innermost radius of the dust torus. According to the model, $r\_$ratio can have the values of 10, 30, 60, 100 and 150.  
 The main dust components are silicate and graphite grains. Silicate grains are responsible for the absorption feature at 9.7 $\mu$m, and $\tau$ is the optical depth at 9.7 $\mu$m. 
 In these models $\tau$ can be  0.1, 0.3, 0.6, 1.0, 2.0, 3.0, 6.0 and 10.0. The dust density distribution is determined by  $r^{\beta} e^{-\gamma |cos\theta |}$. Here, the $\beta$ 
 parameter is related to the radial dust distribution in the torus and can have the values of  -1.00, -0.75, -0.50, -0.25 and 0.00. The $\gamma$ parameter is related to the angular 
 dust distribution in the torus and may have the values of 0.0, 2.0, 4.0 and 6.0. $\theta$ is the opening angle of the torus, it can have the values of 60$^{\circ}$,
 100$^{\circ}$ or 140$^{\circ}$. The angle between equatorial axis and line of sight, $\psi$, can have values between 0.001 and 89.990 in steps of 10 degrees. 
 $\psi = 90^{\circ}$ for Type 1 AGN and $\psi = 0^{\circ}$ for Type 2 AGN.
      
\begin{figure*}
\begin{center}$
\begin{array}{lr}
\includegraphics[scale=0.61]{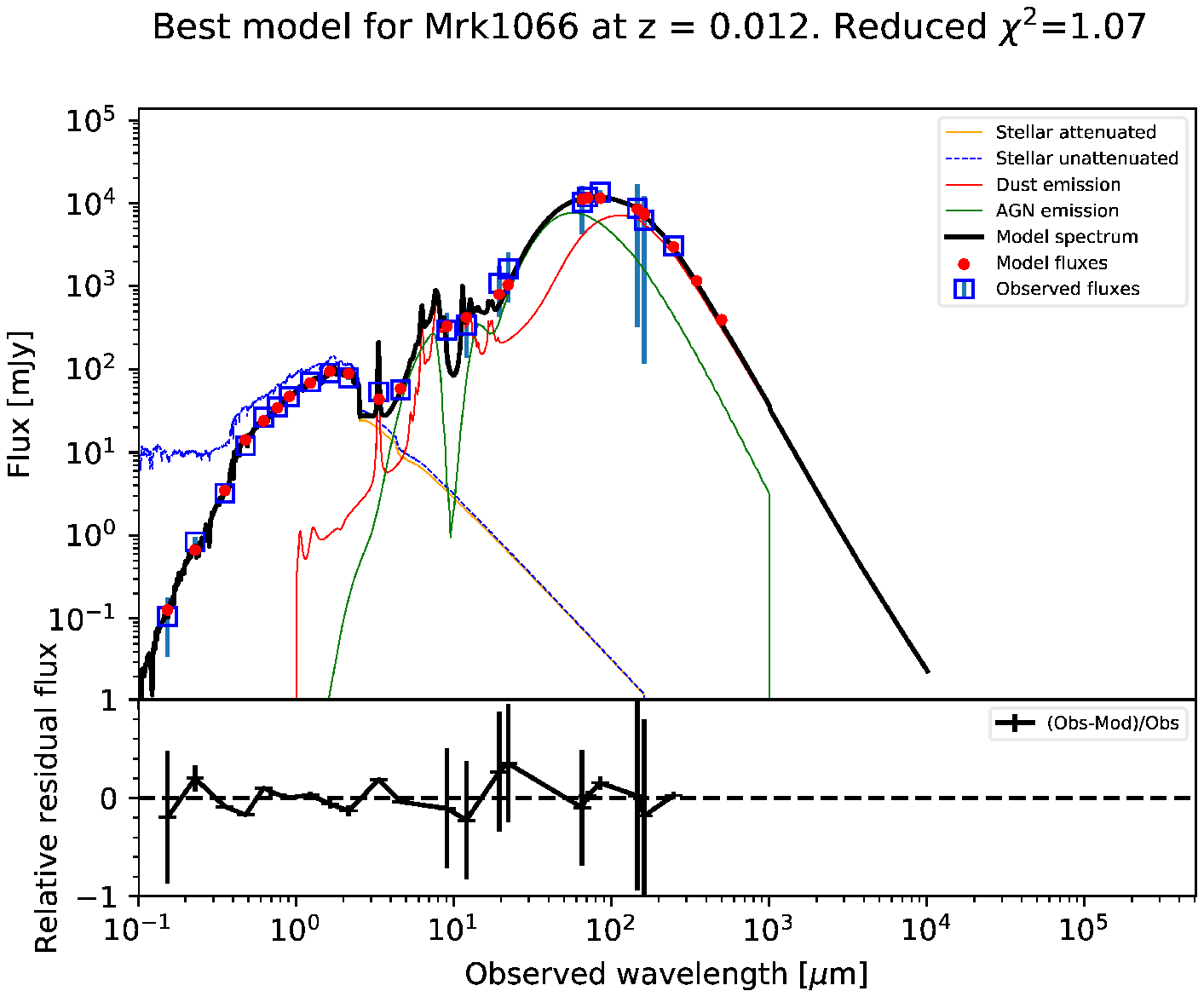}&
\includegraphics[scale=0.61]{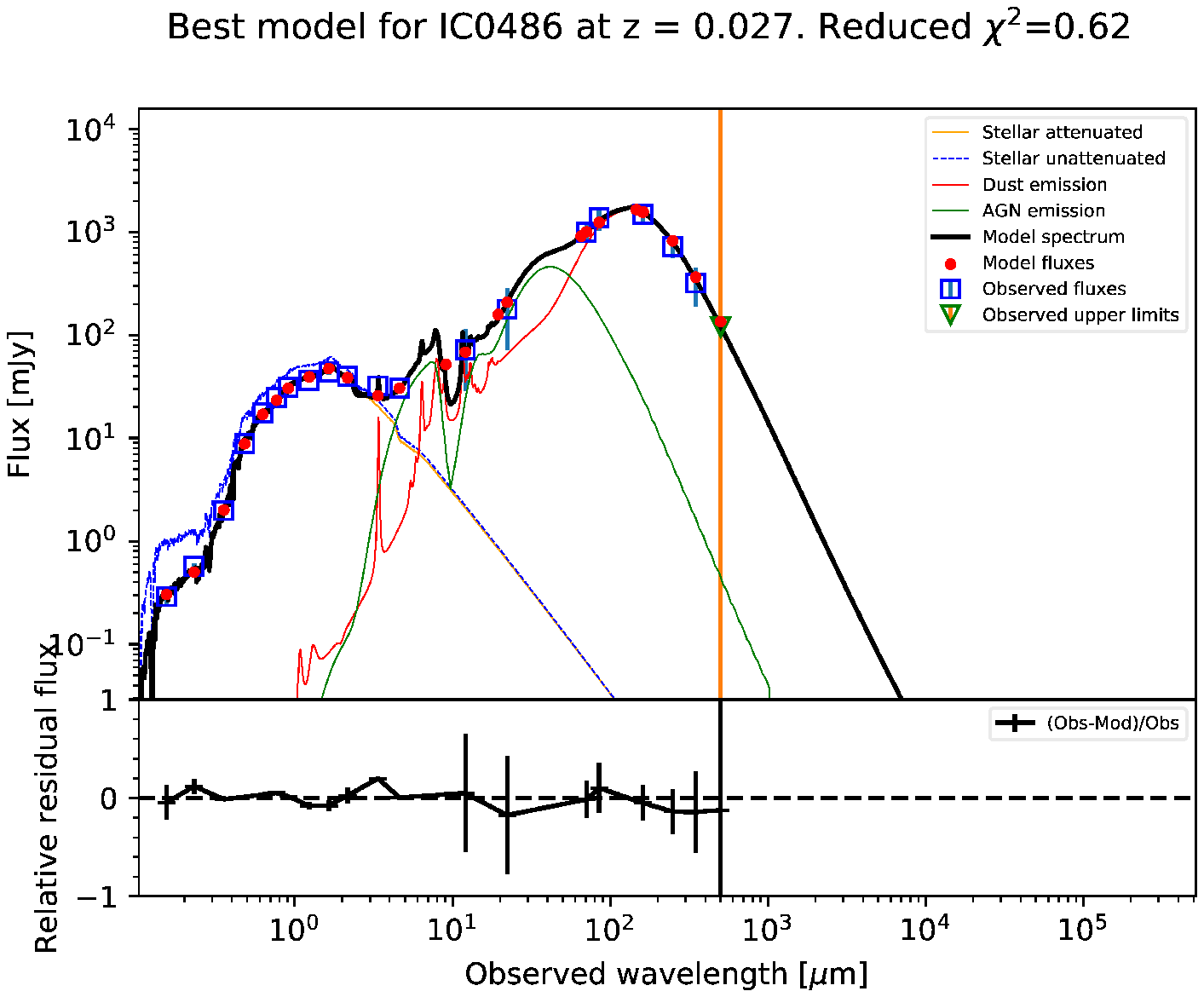}\\ 
\end{array}$
\end{center}
\caption{ Examples of best-fitting models of one CT (left) and one obscured (right) AGN obtained with CIGALE. 
These SEDs are representative of SEDs with full photometric coverage data from UV to FIR. 
Blue squares are the measured flux densities (from \textit{GALEX}, SDSS, 2MASS, \textit{WISE}, \textit{AKARI} and \textit{Herschel}) of the AGN. 
Filled circles (red) are best-fitting model fluxes.
The black solid lines are the best fits given by CIGALE. 
The dashed blue lines, the solid red lines and the solid green lines  show the 
unattenuated stellar emission, the dust emission and the total AGN emission, respectively. 
} 
\label{fig:fig4}
\end{figure*}

In this work, our main interest is to constrain the parameters of the AGN component and measure the AGN and dust luminosities. 
We obtain statistically good fits ($\chi_{reduced}^{2} \leq 5.0$) for most sources except four unobscured AGN (M81, NGC4579, NGC4235, NGC5273). 
We show examples of SED fitting for one CT and one obscured AGN with good photometric coverage from UV to FIR including optical and near-IR bands, in Figure \ref{fig:fig4}. 

We obtain several physical quantities as a result of the SED analysis by CIAGLE. These are:  (i) the host galaxy dust luminosity, $L_{\rm{dust}}$, is measured from the best-fitted dust emission model 
(the red component in Fig. \ref{fig:fig4}) from  the models of \citet{Draine2014}; (ii) the pure AGN luminosity in the IR band, $L\rm{(IR)}_{\rm{AGN}}$ based on the  \citet{Fritz2006} AGN model (the green component in Fig. \ref{fig:fig4}); $L\rm{(IR)}_{\rm{AGN}}$ is the sum of three AGN luminosity components (the direct emission from the central engine, the thermal emission from torus and the scattered emission from the torus); (iii) the total IR 
luminosity $L\rm{(IR)}_{\rm{TOTAL}}$ can be measured as the sum of $L_{\rm{dust}}$ and  $L\rm{(IR)}_{\rm{AGN}}$. The physical origin of $L_{\rm{dust}}$ is the host galaxy dust grains heated by the interstellar radiation \citep{Draine2014}.

Based on the analysis of the 68 unobscured,  65 obscured and 25 CT AGN SEDs we find that 
the fractional AGN emission contribution to the total IR luminosity ($frac_\mathrm{AGN}$) is between 10 and 80\%. 
Based on the $frac_\mathrm{AGN}$ values listed in Table \ref{tab:table1} we identify 7 unobscured, 2 obscured and 18 CT AGN with high ( $\geq$ 40\%) AGN  contribution to the total infrared luminosity. 
These sources  represent IR SEDs with strong AGN luminosity and weak star formation activity from the host galaxy. 
We have also checked if there is a correlation between the $frac_\mathrm{AGN}$ value and the N$_{\rm{H}}$. 
We find that while unobscured and obscured AGN have a similar $frac_\mathrm{AGN}$ distribution between 
10 and 60\%, CT AGN have higher $frac_\mathrm{AGN}$ values within the 10 and 80\% range. Unobscured and obscured AGN do not show a correlation 
 between $frac_\mathrm{AGN}$ and N$_{\rm{H}}$. On the other hand, CT-AGN tend to have higher $frac_\mathrm{AGN}$.

As a result of the SED analysis of 68 unobscured,  65 obscured and 25 CT AGN we find that
while unobscured AGN can have $\psi$ values between 30.10$^{\circ}$ and $89.990^{\circ}$, all obscured and CT AGN have $\psi$ values of  $0.001^{\circ}$or $20.0^{\circ}$ as expected. 
Our analysis confirm that  the angle between equatorial axis and line of sight an important parameter to separate obscured and unobscured AGN based on the SED fitting using this model. 
We find that  CT and obscured AGN have $\theta$ values of  140$^{\circ}$ (52 per cent), 100$^{\circ}$ (32 per cent) or 60$^{\circ}$ (15 per cent). 
Unobscured AGN can have $\theta$ values of  140$^{\circ}$ (53 per cent), 100$^{\circ}$ (32 per cent) or 60$^{\circ}$ (14 per cent). 
We find that unobscured, obscured and CT  AGN mostly have $\beta = -0.5$ or $\beta = -1.00$. 
Unobscured, obscured  and CT AGN mostly have $\gamma=4.0$ and $\gamma=6.0$. 
Models with $r\_ratio$ equal to 150 are  favoured for unobscured, obscured and CT AGN. 
Models with $\tau$ equal to 10.0 provide good fits for the most of the unobscured, obscured and CT AGN. 

Our SED analysis with CIGALE shows that obscured and CT AGN can be identified based on the  parameters of the \citet{Fritz2006} model, the most critical parameters are 
the angle between equatorial axis and line of sight $\psi$, angular opening angle of the torus, $\theta$ . 
Once the best fitted SED given by CIGALE has $\psi = 0.001^{\circ}$ or $\psi = 10.0^{\circ}$, and  $\theta= 140^{\circ}$, then 
it is very likely that the source is an obscured/CT AGN. 

\begin{table*}
           \centering
           \caption{
           Parameters used in the SED fitting by CIGALE.
	}
           \label{tab:table2}
            \begin{tabular}{lc}
           \hline
            Parameter  & Value\\
            \hline
                              &SFH \\
            \hline
            Main stellar age      & 200,300,400,500,600,700,800,900,1000, 2000,2600,2900,3000,3500,4000,5000,10000\\
            burst age                &  1,2,6,10, 20,30,40,50,80,90,100,150,200,380,400,450,600,900,950,1400\\
            $\tau_{main}$         &  20,100,200,500,1000,1200,1300,3000,6000,7000\\
            $\tau_{burst}$        &  1,5,10, 50\\
            \hline
                              & Dust emission \\
           \hline
            q$_{PAH}$  &  0.47, 1.12, 1.77, 2.50\\
            U$_{min}$   &  0.170, 0.200, 0.250, 0.350, 0.500, 0.600, 0.700,1.000, 1.200, 1.500, 1.700, 2.000, 2.500, 3.000,\\
                                &  3.500, 4.000, 5.000, 6.000, 7.000, 8.000, 10.00, 12.00\\
            $\alpha$      & 1.7, 1.8, 1.9, 2.0, 2.1, 2.2, 2.3, 2.4, 2.6, 2.7, 2.8, 2.9, 3.0 \\
            $\gamma$   &  0.02, 0.1, 0.2, 0.3, 0.4\\
		\hline
		               & Dust attenuation \\
		\hline
		slope ISM   &  -0.7,-0.5,-0.3,-0.1\\
		slope birth clouds & -1.3, -0.7\\
		\hline
		               & AGN \\
		\hline
		$R_{max}/R_{min}$    & 10, 30, 60, 100, 150\\
		$\tau$                          & 0.1, 0.3, 0.6, 1.0, 2.0, 3.0, 6.0, 10.0.\\
		$\beta$                        & -1.00, -0.75, -0.50, -0.25, 0.00\\
		$\gamma$                   & 0.0, 2.0, 4.0, 6.0\\
		$\theta$                         & 60, 100, 140 \\
		$\psi$                          & 0.001, 10.100\\
		$frac_{AGN} $             &  0.1, 0.2, 0.3, 0.4, 0.5, 0.6, 0.7\\
		\hline
	\end{tabular}
\end{table*}

\begin{table*}
           \centering
           \caption{
           Photometric filters used in the SEDs and the detection rate out of 338 AGN in our sample at each band. Columns: (1) Telescope/survey name. 
           (2) Filter name. (3) Efficient wavelength of the filter. (4) Number of detections of the 338 AGN in our sample at each band.  
           (5) Number of sources that has at least one detection at each region: region A is UV-optical (covers FUV, NUV,u,g,r,i,z bands), B is NIR (covers J, H,Ks bands), C is MIR (covers w1,w2,w3,w4,S9W,L18W bands) and D is FIR (covers N60,WIDE-S,WIDE-L,N160,PACS blue,PACS red bands ). Each region is separated by the horizontal lines.
	}
           \label{tab:table3}
            \begin{tabular}{lcccc}
           \hline
           \hline
            Telescope/  & Filter  & $\lambda_\mathrm{eff}$($\mu$m) &Detection & Number of sources  \\
              Survey      &          &                                           & rate at   & at least one detection \\
                                &        &                                            &each band &in each region \\
            \hline
 \textit{GALEX}  & FUV   		 & 0.15 & 173  & \\
                          & NUV  		 & 0.23 & 222 & \\
            SDSS    & u        		 & 0.36  &75 & \\
                          & g        		 & 0.46 &75 & A= 244\\
                          & r         		 & 0.61 &75 & \\
                          & i         		& 0.74 &75 & \\
                          & z        		& 0.89 &75 & \\
            \hline              
           2MASS  & J        		& 1.24 & 335 & \\
                          & H       		& 1.66 & 335 & B=335 \\
                          &Ks      		& 2.16 & 335 & \\
           \hline               
  \textit{WISE}    & w1      & 2.25 & 256 & \\
                          & w2      & 4.60  & 256 & \\
                          & w3      & 11.56  & 255 & \\
                          & w4      & 22.09  & 255 &  C=305 \\
   \textit{AKARI}  & S9W    & 9 & 149 & \\
                          & L18W  & 18 & 186 & \\
          \hline                
    \textit{AKARI}  & N60     & 65 & 92 & \\
                          & WIDE-S   & 90 & 333 & \\
                          & WIDE-L   & 140 & 120 & \\
                          & N160        & 160 & 23 & D=338 \\
 \textit{Herschel}  & PACS blue  &   68.927  & 155 & \\
                           & PACS red  & 153.95 & 155 & \\
         \hline                  
 \textit{Herschel}  &PSW  & 242.82 & 187& \\
                           &PMW  & 340.89 & 187 & \\
                           &PLW   & 482.25  &  187 & \\    
		\hline
	\end{tabular}
\end{table*}

 \section{Results and Discussion} \label{S:results} 
 
 \subsection{Average SEDs of Unobscured Obscured and Compton-thick AGN} \label{S:avseds}
 
\begin{figure*}
\begin{center}$
\begin{array}{cc}
\includegraphics[]{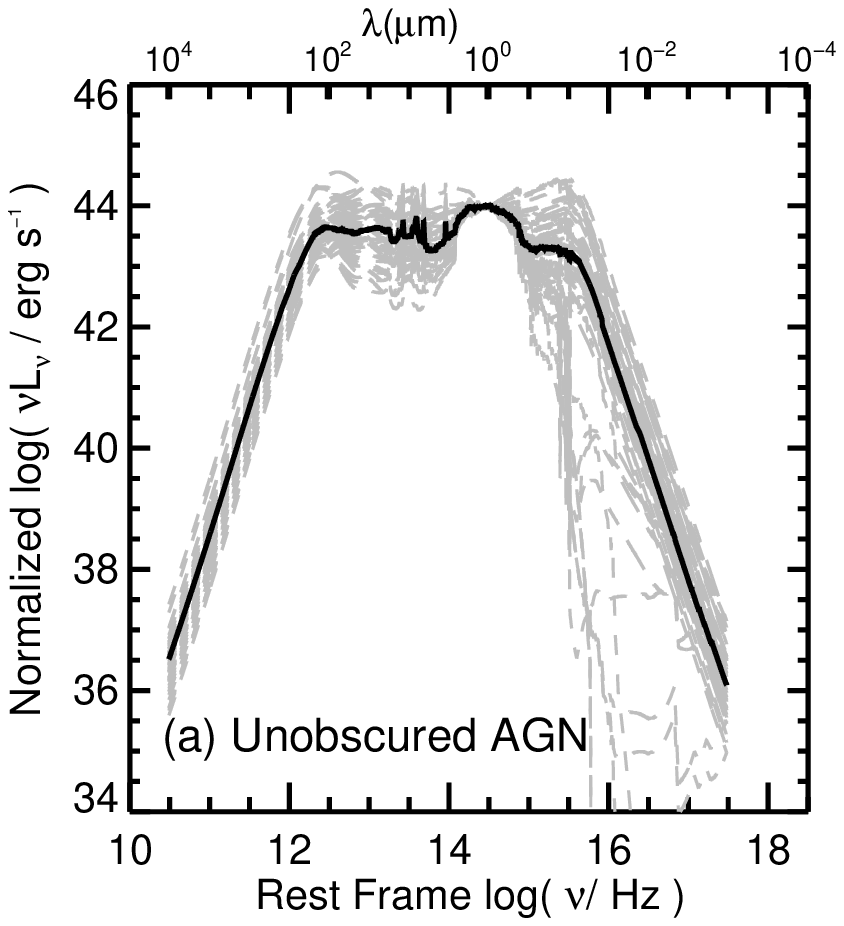} &
\includegraphics[]{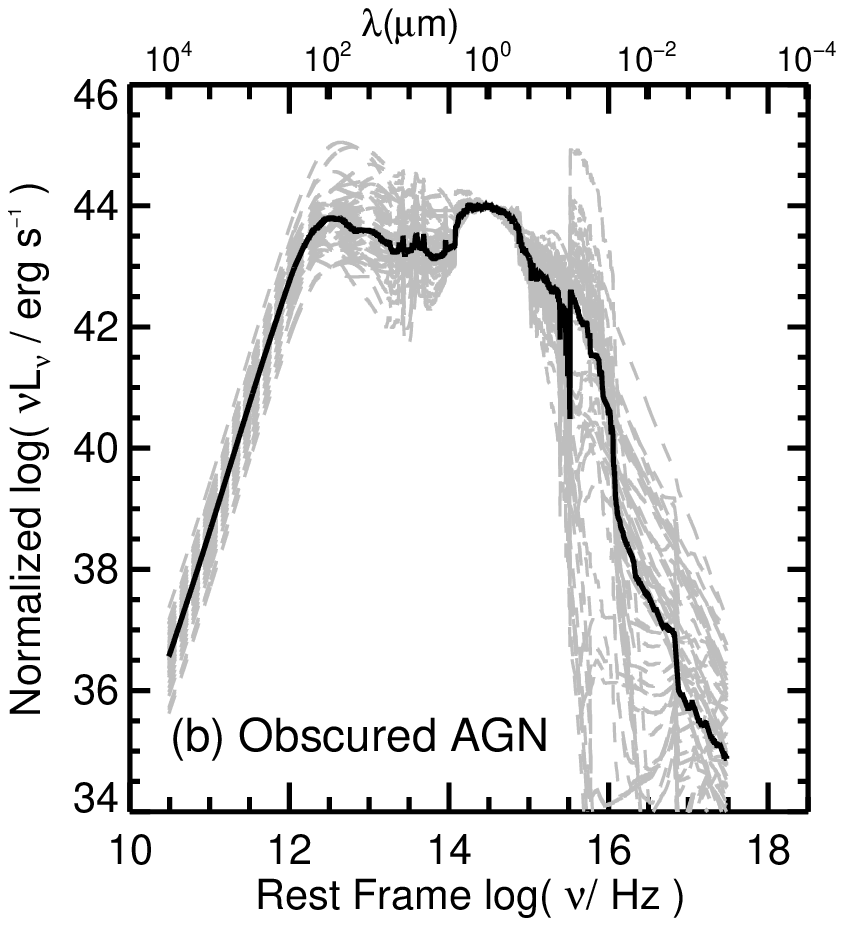} \\
\includegraphics[]{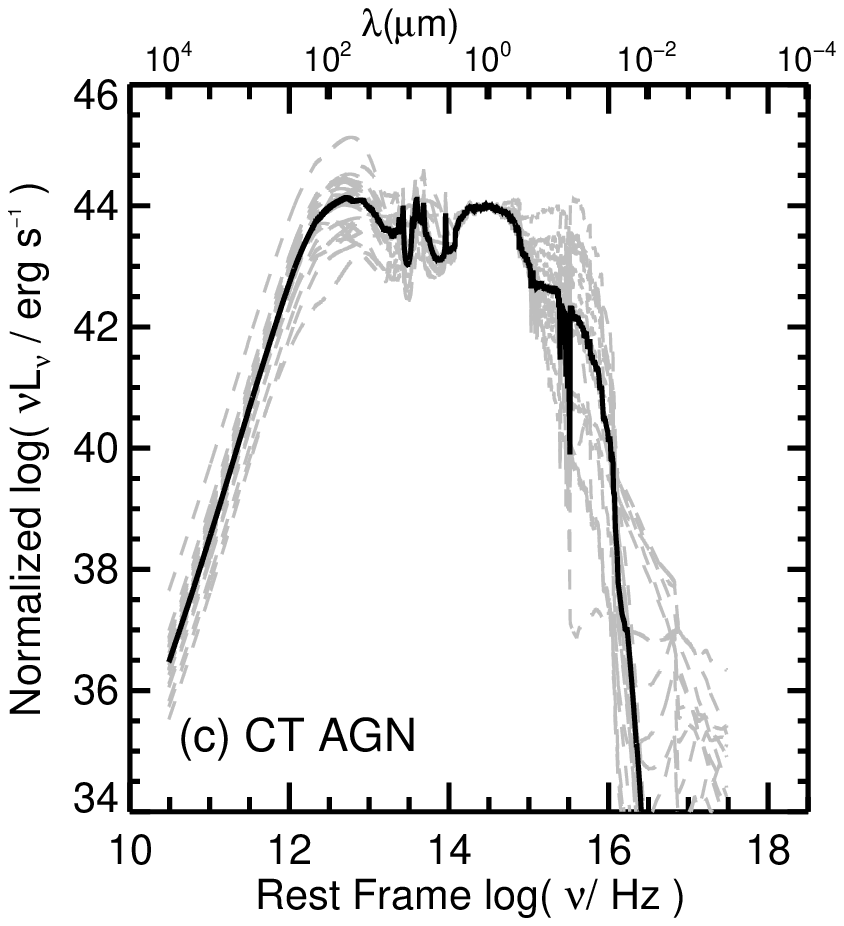} &
\includegraphics[]{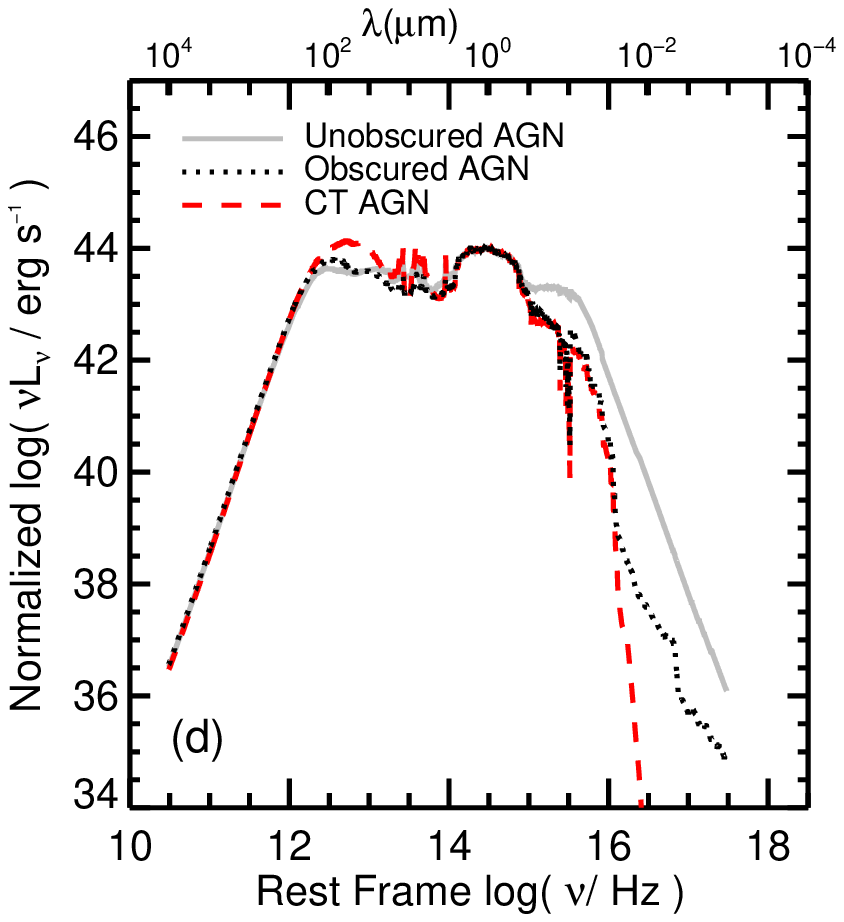} \\
\end{array}$
\end{center}
\caption{ Obtained mean SEDs of unobscured (panel a), obscured (panel b) and CT (panel c) AGN populations as a result of SED analysis by CIGALE. In panel (d) the mean SEDs are shown on top of each other. 
All SEDs are normalised at rest frame 1$\mu$m. 
} 
\label{fig:fig5}
\end{figure*}

 As a result of our SED analysis across the UV to far-IR wavelength region, we make average SEDs of unobscured, obscured and CT AGN. We build the mean best-fitted SEDs by CIGALE as follows: 
(i) the fluxes densities at each frequency are converted to luminosity; (ii) we calculate the mean SED at each $\log\nu$ grid point ($\Delta\log\nu = 0.0007$), we normalise each SED at 1$\mu$m. 
Fig. \ref{fig:fig5} shows the mean SEDs of unobscured, obscured and CT AGN in panels (a), (b) and (c), respectively. Panel (d) shows the mean SEDs of each population together. 
As seen in panel (d) in the optical-UV region, unobscured AGN have strong contribution compared to the obscured/CT AGN. This is consistent with the adopted AGN model of \citet{Fritz2006} since 
the emission from the inner the torus is completely absorbed due to the optically thick torus. Therefore, the mean SEDs of the obscured/CT AGN dominated by the stellar population emission in the optical-UV region. 
Near rest frame 10$\mu$m the mean  AGN continuum is slightly stronger for unobscured AGN compared to the that of obscured/CT AGN. 
This can be understood as a result of the relatively stronger silicate absorption feature seen in the obscured/CT AGN. 
The mean SEDs are very similar in between the 1 - 20$\mu$m mid-IR range. The obtained mean SED properties of obscured, unobscured and CT AGN in the optical to mid-IR range agree  
with previously obtained composite SEDs of Type 1 and Type 2 \citep[e.g.,][]{Hickox2017}. 
In the MIR region PAH emission from host-galaxy star-light is stronger in the mean SED of the CT AGN. 
The mean SED of the CT AGN shows a stronger FIR bump compared to the un/obscured mean SEDs. This is consistent with the more FIR emission observed in type-2 quasars \citep[e.g.,][]{Hiner2009,Chen2015}. 
The far-IR emission is expected to be formed by the cold-dust emission heated by star formation. However, as suggested by  \citet{Hiner2009} dust heated by AGN can have a significant contribution to the far-IR emission. 
 The stronger FIR emission of CT AGN supports a connection between the AGN obscuration and host galaxy dust emission.

\subsection{Relations Between the Infrared Luminosities and Ultra Hard X-ray Luminosity} \label{S:LxLir}

\begin{figure*}
 \begin{center}$
 \begin{array}{lll}
  \includegraphics[scale=0.66]{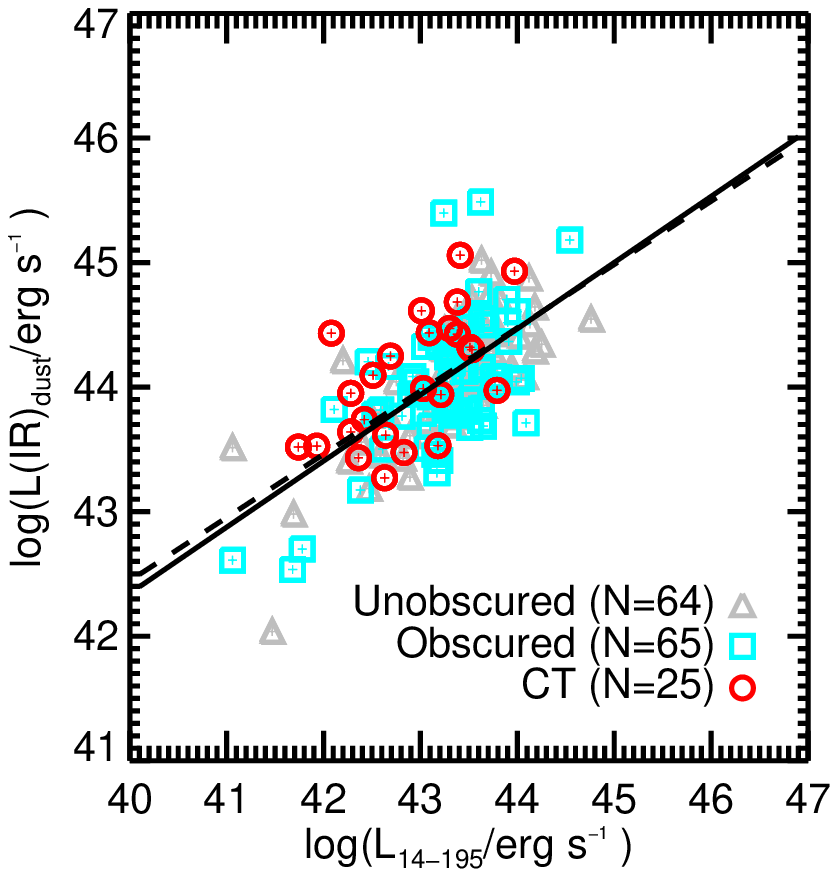}& 
 \includegraphics[scale=0.66]{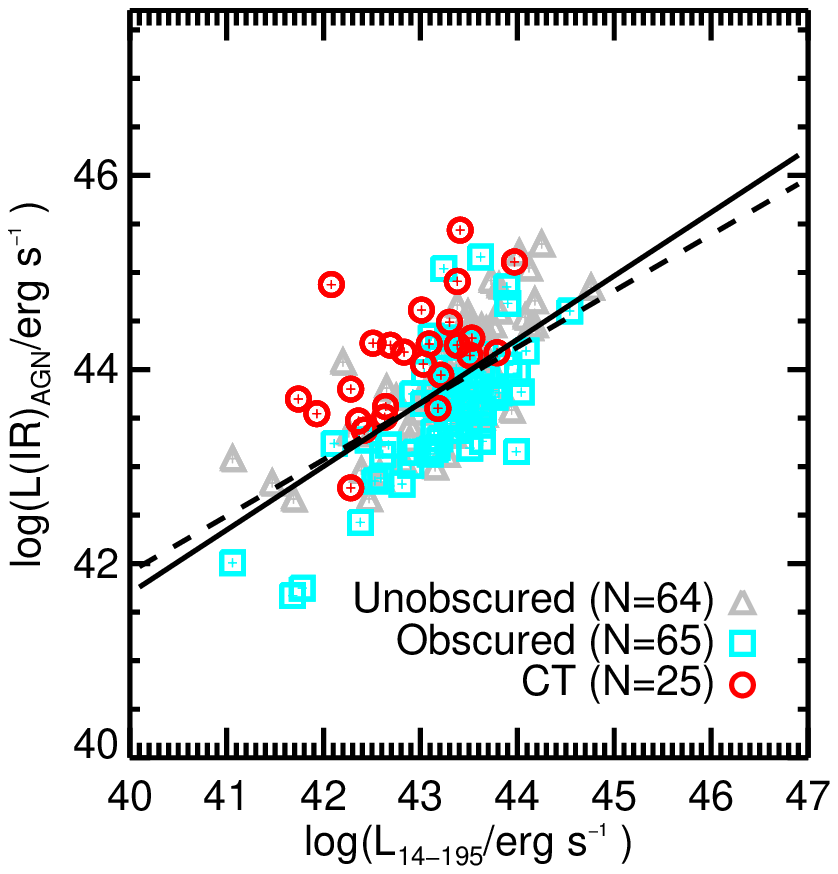} & 
 \includegraphics[scale=0.66]{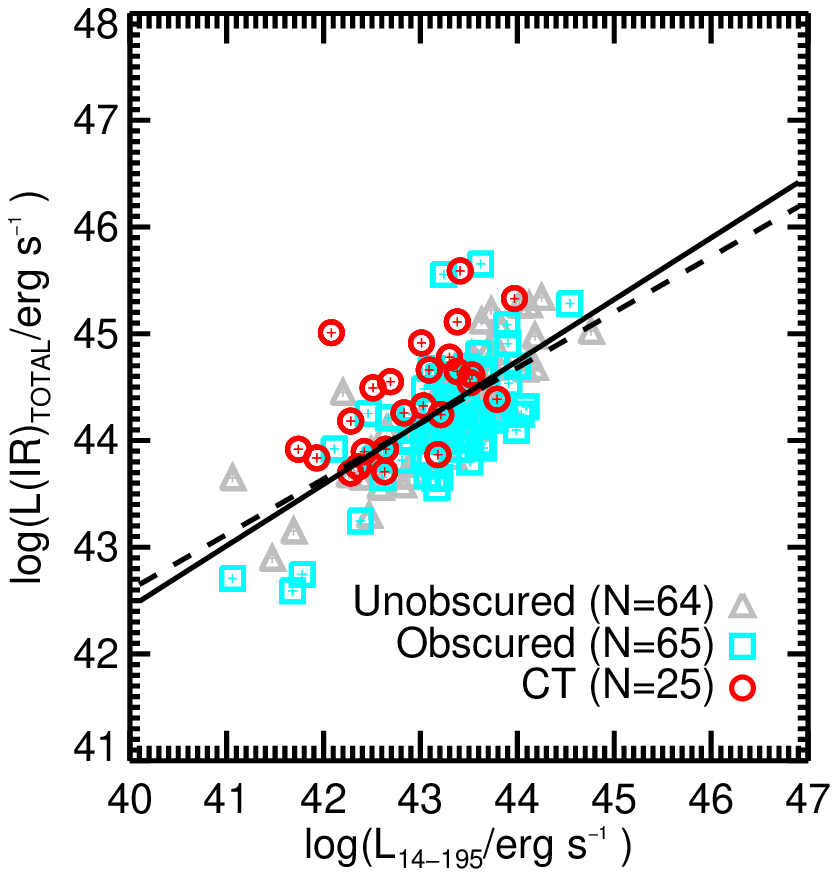} \\
 \end{array}$
 \end{center}
 \caption{ Relations between the  14-195\,keV hard X-ray luminosity and $L_{\rm{dust}}$ (left panel), the pure AGN luminosity in the IR band, $L\rm{(IR)}_{\rm{AGN}}$ (middle panel) and  
the total IR luminosity (right panel) $L\rm{(IR)}_{\rm{TOTAL}}$. Triangles, squares and the circles represent the  unobscured, obscured and CT AGN, respectively. 
The black solid lines show the best-fitting relationships for all sources. The dashed lines represent the best-fitting relationships for only obscured and CT AGN.} 
 \label{fig:fig6}
 \end{figure*}

As a result of the SED fitting analysis in \S \ref{S:seds}, we measure the dust luminosity, $L_{\rm{dust}}$, the pure AGN luminosity in the IR band, $L\rm{(IR)}_{\rm{AGN}}$ and the total IR 
luminosity $L\rm{(IR)}_{\rm{TOTAL}}$ as the sum of the $L_{\rm{dust}}$ and  $L\rm{(IR)}_{\rm{AGN}}$. 
Here we quantify the relationships between $L_{\rm{dust}}$, $L\rm{(IR)}_{\rm{AGN}}$, $L\rm{(IR)}_{\rm{TOTAL}}$ and the ultra hard X-ray luminosity  in the 14-195\,keV band. 
Figure \ref{fig:fig6} shows the relations between $L_{\rm{dust}}$, $L\rm{(IR)}_{\rm{AGN}}$, $L\rm{(IR)}_{\rm{TOTAL}}$ and $L_{\rm{14 - 195}}$ for unobscured, obscured and CT AGN in our sample. 
Clear, positive correlations are seen for $L_{\rm{dust}}$-$L_{\rm{14 - 195}}$ (left panel), $L\rm{(IR)}_{\rm{AGN}}$-$L_{\rm{14 - 195}}$ (middle panel) and $L\rm{(IR)}_{\rm{AGN}}$-$L_{\rm{14 - 195}}$ (right panel). 
We determine the significance of the correlations using Pearson correlation coefficient \citep[$r$,][]{Pearson1895}. 
We measure $\sim r=0.6$ with a p-value equal to zero which indicates a moderately strong correlation for all luminosities.  

To characterise the $L_{dust}$-$L_{\rm{14 - 195}}$, $L\rm{(IR)}_{\rm{AGN}}$-$L_{\rm{14 - 195}}$ and  $L\rm{(IR)}_{\rm{TOTAL}}$-$L_{\rm{14 - 195}}$ relationships, we adopt the following parameterisation:
 \begin{equation}\label{Eq:1}
\log(L_{dust})=A_{dust}+\alpha_{dust} \log(L_{\rm{14 - 195}}) +\epsilon_{dust}.
 \end{equation}
 
\begin{equation}\label{Eq:2}
\log(L\rm{(IR)}_{\rm{AGN}})=A_{AGN}+\alpha_{AGN} \log(L_{\rm{14 - 195}}) +\epsilon_{AGN}.
 \end{equation}
 
 \begin{equation}\label{Eq:3}
\log(L\rm{(IR)}_{\rm{TOTAL}})=A_{TOTAL}+\alpha_{TOTAL} \log(L_{\rm{14 - 195}}) +\epsilon_{TOTAL}.
 \end{equation}
 
where A is the zero point, $\alpha$ is the slope and $\epsilon$ is the estimated scatter. 
We establish the best-fitting relationships for both luminosities by the Bayesian regression method of \citep{Kelly2007} that accounts for scatter ($\epsilon$) and computes the posterior probability 
distributions of the parameters.

For the $L_{dust}$-$L_{\rm{14 - 195}}$ relationship the best-fitting parameters are: $A_{dust}$= 22.08$\pm$3.27, $\alpha_{dust}$= 0.51$\pm$0.07 and $\epsilon_{dust}$= 0.43$\pm$0.03. 
The best-fitting relationship for all sources is shown as the solid line. The dashed line, which is very similar to the solid line, shows the best-fitting relationship of obscured and CT AGN. 
Best-fitting parameters for $L\rm{(IR)}_{\rm{AGN}}$-$L_{\rm{14 - 195}}$ relationship are: 
$A_{AGN}$= 15.50$\pm$2.91, $\alpha_{AGN}$= 0.66$\pm$0.07 and $\epsilon_{AGN}$= 0.53$\pm$0.03. The best-fitting relationship is shown as the solid black line for the unobscured, obscured and CT AGN. 
When we only consider the obscured and CT AGN we find a slightly shallower slope of $\alpha_{AGN}$= 0.58$\pm$0.10 as shown by the dashed black line. 
Best-fitting parameters for $L\rm{(IR)}_{\rm{TOTAL}}$-$L_{\rm{14 - 195}}$ relationship are: 
$A_{TOTAL}$= 19.35$\pm$2.23, $\alpha_{TOTAL}$= 0.57$\pm$0.05 and $\epsilon_{TOTAL}$= 0.40$\pm$0.02. 
The best-fitting relationship (dashed line) for obscured and CT AGN is slightly shallower than that of the all  sources (solid line). 
The slope of the dashed line is $\alpha_{TOTAL}$= 0.52$\pm$0.08. 
The obtained scatter values around the best-fitting relationships are similar (between 0.4 and 0.5) for all panels.  
The  $L_{\rm{dust}}$-$L_{\rm{14 - 195}}$ relationship slope is shallower compared to that of $L\rm{(IR)}_{\rm{AGN}}$-$L_{\rm{14 - 195}}$.
Since $L\rm{(IR)}_{\rm{AGN}}$ is the pure AGN luminosity and $L_{\rm{dust}}$ originates from starburst radiation we expect to have different slopes for these relationships. 

The relationship between the AGN mid-IR luminosity  and X-ray luminosity has been investigated by previous studies \citep[e.g.,][]{Asmus2015,Ichikawa2017,Ichikawa2019}.
\citet{Ichikawa2019} decomposed the IR SED of the 587 AGN in the 70-month \textit{Swift}/BAT sample into starburst and AGN components. 
They quantified the relationship between the 12$\mu$m, mid-IR luminosities and the ultra hard X-ray luminosity  in the 14-195\,keV for \textit{Swift}/BAT 70-month AGN. 
When we compare the $L\rm{(IR)}_{\rm{AGN}}$-$L_{\rm{14 - 195}}$ relationship slope with the mid-IR AGN luminosity $L_{\rm{14 - 195}}$ relationship slope values between 0.96-1.06 obtained by
 \citet{Asmus2015,Ichikawa2017,Ichikawa2019}, we obtain a much shallower slope. The obtained shallower slope in this work might be related to the sample difference. 
 First of all our AGN sample (158) is much smaller compared to that of \citet{Ichikawa2017,Ichikawa2019} and therefore it does not expand to low and high luminosity ranges in both axis. 
 Especially, our sample do not have any AGN with $\log L_{\rm{14 - 195}} > 45.0$ and $L\rm{(IR)}_{\rm{AGN}} < 41.5$, therefore we obtain a shallower slope compared to other studies 
\citep[e.g.,][]{Asmus2015,Ichikawa2017,Ichikawa2019} whose samples include higher $L_{\rm{14 - 195}}$ and lower  $L\rm{(IR)}_{\rm{AGN}}$ objects. 
An other difference that may result in different slope values is the difference between the measured IR luminosities, while we have the AGN luminosity in the total IR range previous works 
\citep[e.g.,][]{Asmus2015,Ichikawa2017,Ichikawa2019} only consider the mid-IR luminosity. We also note that, slope values of $\sim$0.65 \citep[e.g.][]{Netzer2009,MatsuokaWoo2015} and $\sim0.8$  \citep{Ichikawa2017} 
are obtained for the relationship between the far-IR  
luminosity and the bolometric AGN luminosity. The obtained $\alpha_{dust}$= 0.51$\pm$0.07 slope in this work is close to the slope values obtained in these previous studies. 

\begin{figure*}
 \begin{center}$
 \begin{array}{lll}
  \includegraphics[scale=0.9]{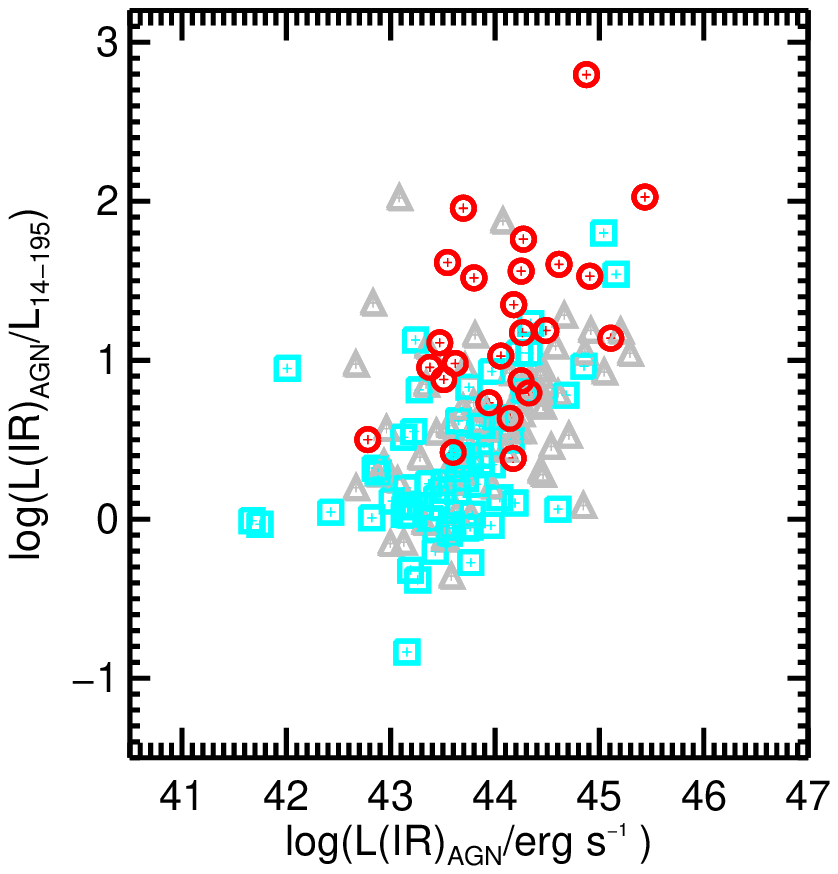}& 
 \includegraphics[scale=0.9]{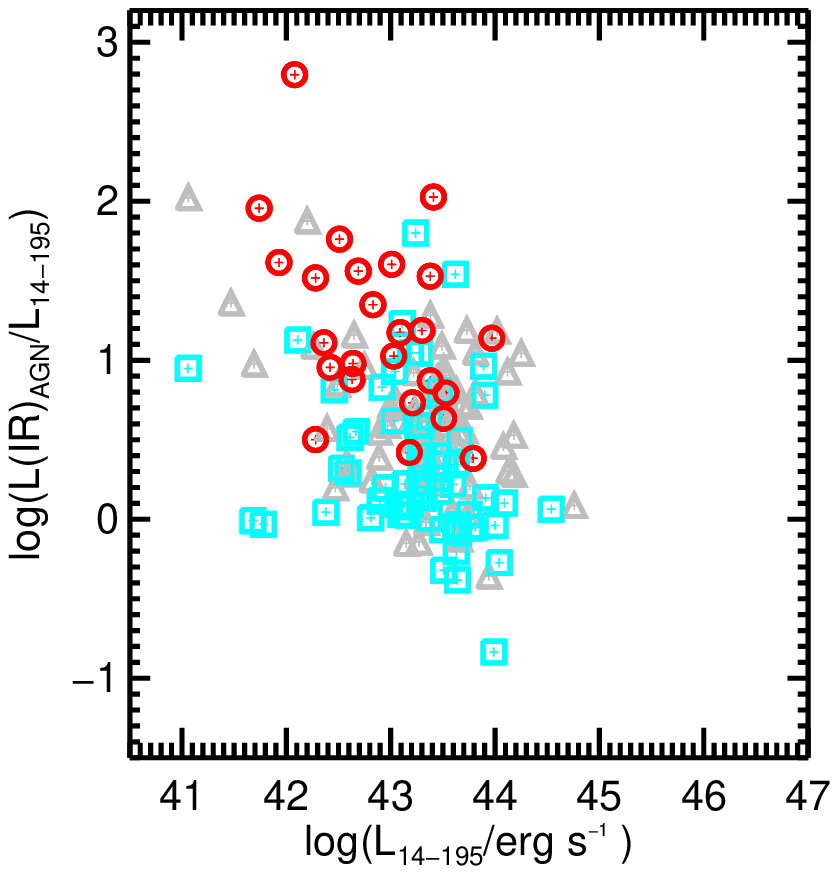} \\ 
 \end{array}$
 \end{center}
 \caption{ $L\rm{(IR)}_{\rm{AGN}}$ to $L_{\rm{14 - 195}}$  fraction as a function of $L\rm{(IR)}_{\rm{AGN}}$ (left panel) and 
 $L_{\rm{14 - 195}}$ (right panel). See Fig. \ref{fig:fig6} for the symbol code.} 
 \label{fig:fig7}
 \end{figure*}

 As shown in Figure \ref{fig:fig7} we also check the fraction of the pure AGN luminosity  to the hard X-ray luminosity as a  function $L\rm{(IR)}_{\rm{AGN}}$ (left panel) and 
 $L_{\rm{14 - 195}}$ (right panel). As seen in both panels,  pure AGN luminosity  to the hard X-ray luminosity fraction of obscured and CT AGN do not show a clear separation 
 from the unobscured AGN. 
 
\begin{figure}
 \begin{center}$
 \begin{array}{c}
  \includegraphics[]{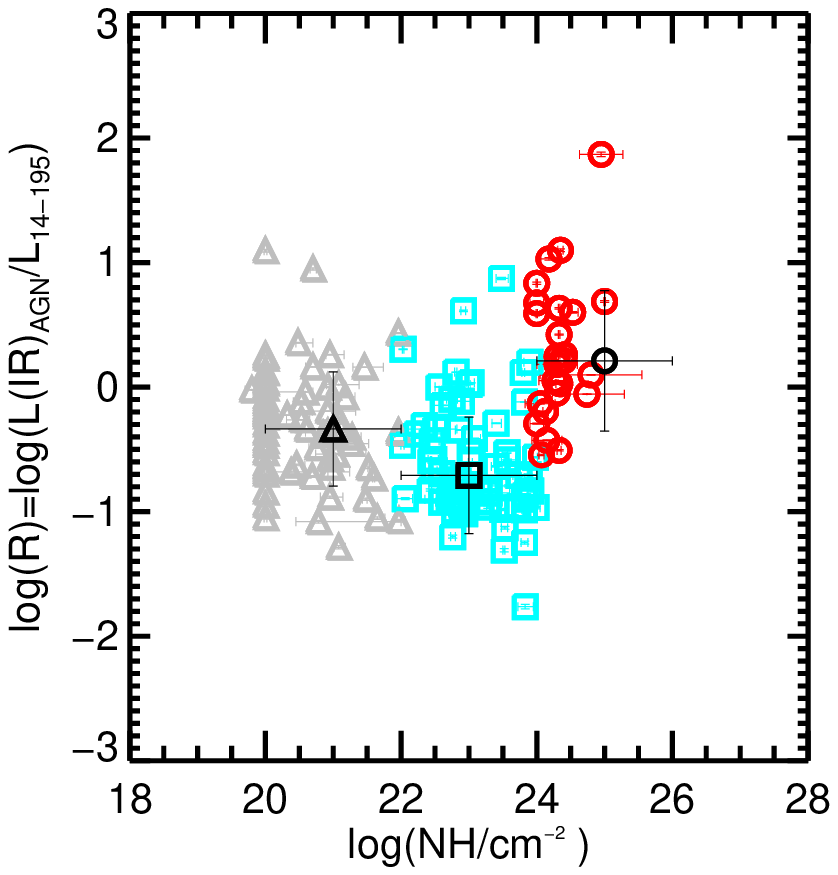}
 \end{array}$
 \end{center}
 \caption{ $R= L\rm{(IR)}_{\rm{AGN}}$/$L\rm{(BOL)}_{\rm{AGN}}$  versus N$_{\rm{H}}$. See Fig. \ref{fig:fig6} for the symbol code. 
 The black symbols represent the median $R$ values in each N$_{\rm{H}}$ bin with the x-axis error bar showing the range of the N$_{\rm{H}}$ bin and the
 y-axis error bar showing the inter-percentage range with 68.2\% of the unobscured, obscured and GT AGN samples.} 
 \label{fig:fig8}
 \end{figure}

 The measured $L\rm{(IR)}_{\rm{AGN}}$ as a result of the SED fitting can be used to calculate $R= L\rm{(IR)}_{\rm{AGN}}$/$L\rm{(BOL)}_{\rm{AGN}}$, 
 which is a good indicator of dust covering factor  \citep[e.g.][]{Elitzur2012}. For comparison we follow  \citet{Ichikawa2019} and use a constant bolometric factor of
  8.47 \citep[][]{Ricci2017a} to obtain $L\rm{(BOL)}_{\rm{AGN}}$. In Figure \ref{fig:fig8} we show $R$ versus N$_{\rm{H}}$ in order to see the differences of covering factors among unobscured, obscured and CT AGN. 
  We find that the median covering factor of the CT AGN ($\log(R_{\rm{ct}})$=0.21) is larger than the median covering factor of the obscured 
  ($\log(R_{\rm{obscured}})$=-0.71) and  the median covering factor of the unobscured ($\log(R_{\rm{unobscured}}$=-0.34) AGN. For our sample we see that the median covering factor of the unobscured 
  AGN is larger than that of the obscured AGN, however we note that the most majority of the obscured and unobscured AGN have $R$ values in a similar range. For a larger sample of \textit{Swift}/BAT 
  AGN \citet{Ichikawa2019} found that obscured AGN have a larger covering factor compared to  unobscured AGN. The found higher covering factor values of the CT AGN is consistent with the result of \citet{Ichikawa2019}. 

 \subsection{Infrared colours and luminosities versus N$_{\rm{H}}$} \label{S:colvsnh}
 
\begin{figure*}
 \begin{center}$
 \begin{array}{ll}
  \includegraphics[]{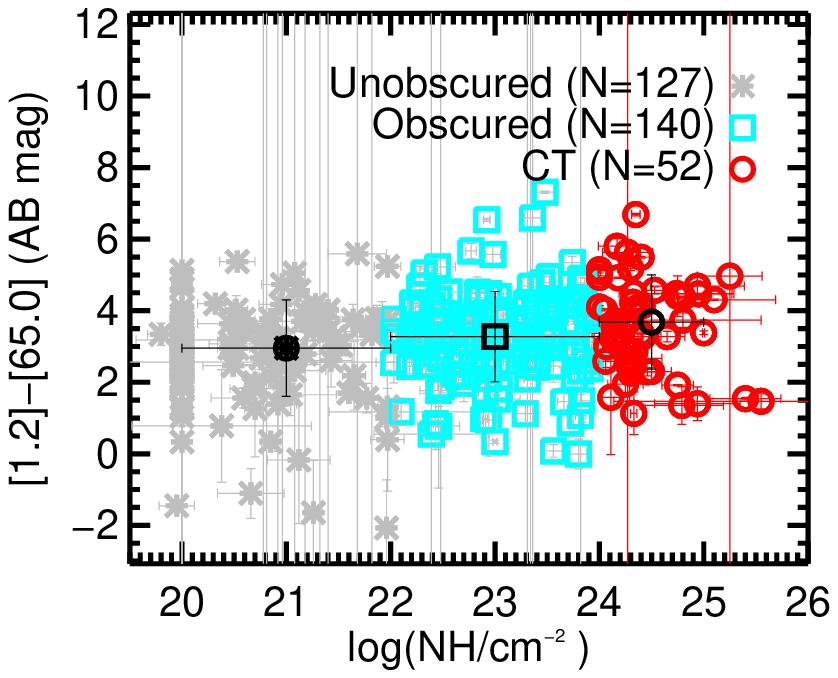} & 
 \includegraphics[]{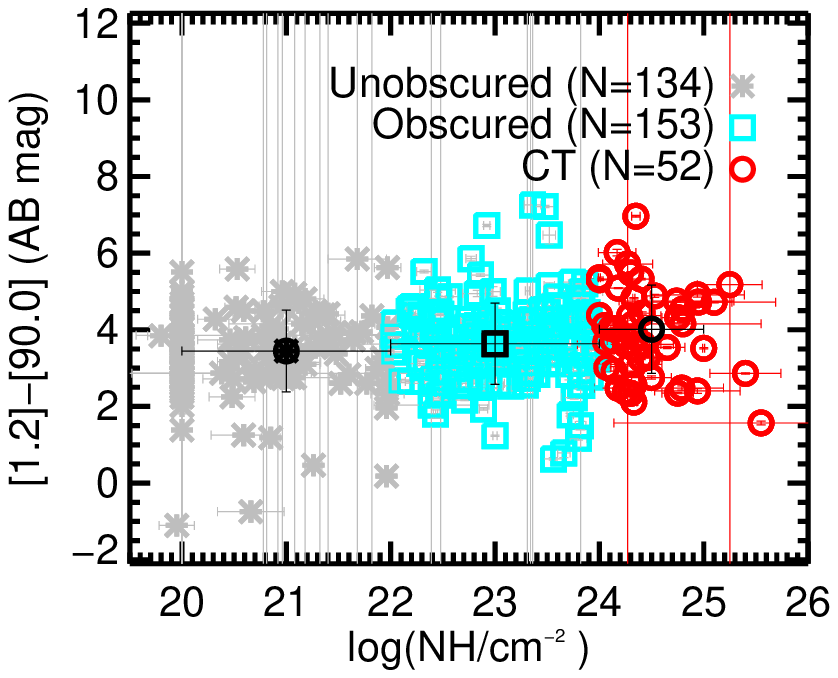} \\
  \includegraphics[]{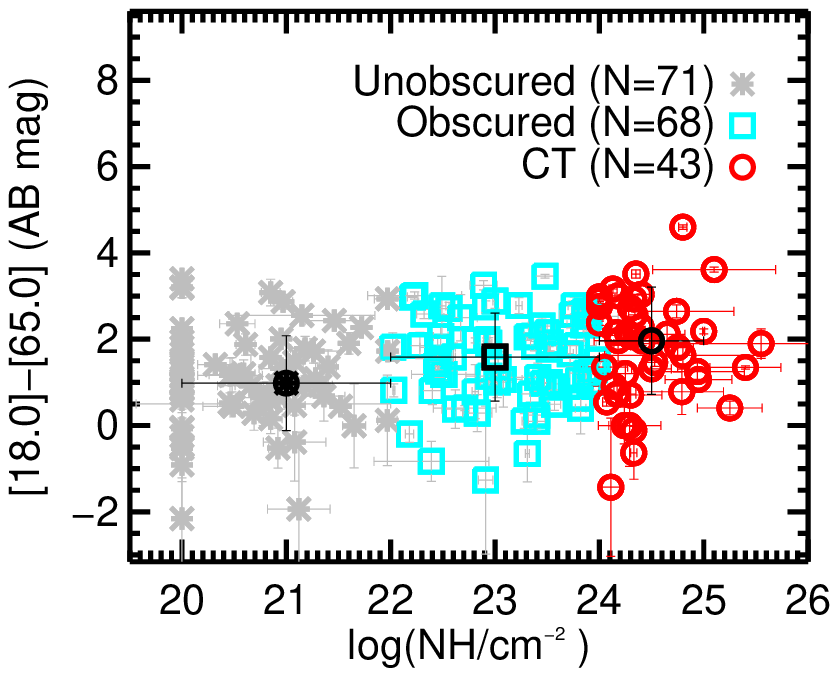} & 
 \includegraphics[]{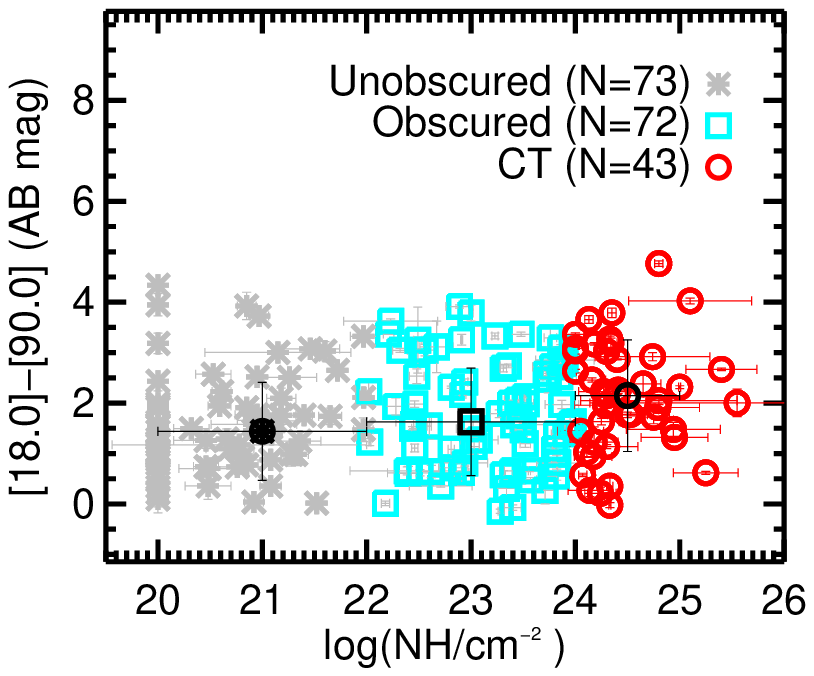} \\
   \includegraphics[]{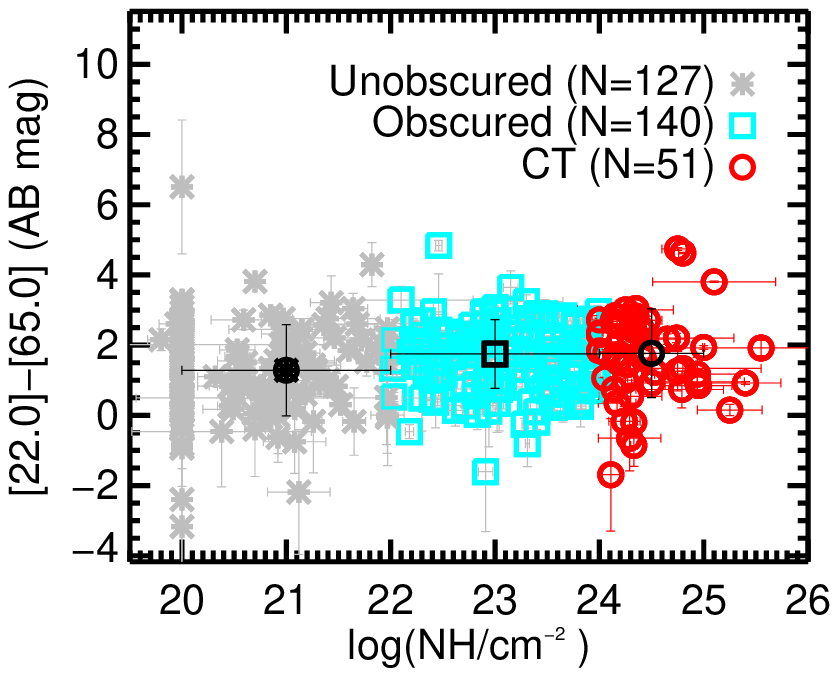} & 
 \includegraphics[]{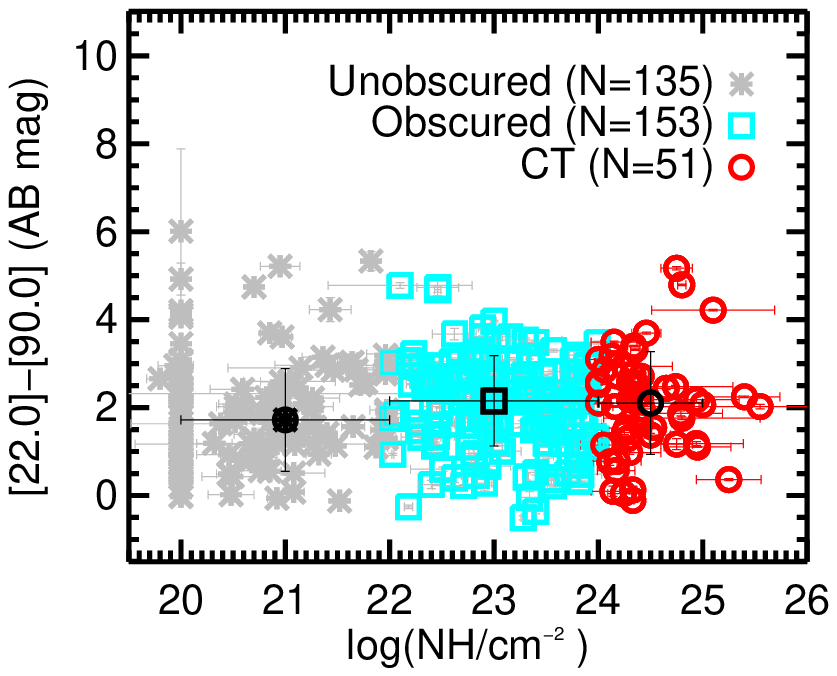} \\
 \end{array}$
 \end{center}
 \caption{ Infrared colour dependence on N$_{\rm{H}}$. The black symbols represent the median colours in each N$_{\rm{H}}$ bin. 
 Median 1.25$\mu$m - 65 $\mu$m, 1.25$\mu$m - 90 $\mu$m, 18 $\mu$m - 65 $\mu$m, 18 $\mu$m - 90 $\mu$m, 22 $\mu$m - 65 $\mu$m, 22 $\mu$m - 90 $\mu$m colours show an increasing trend with N$_{\rm{H}}$.} 
 \label{fig:fig9}
 \end{figure*}

Here we investigate if the IR colours depend on the N$_{\rm{H}}$. 
Near-IR (NIR) photometry is a good tracer of direct-stellar component, the MIR photometry is a good tracer of AGN torus and the FIR photometry is a good tracer of host galaxy dust emission. 
Therefore, NIR-FIR and MIR-FIR colours have physically different origins. NIR-FIR colours represent the energy balance between the direct-stellar versus dust emission from the host galaxy, 
while the MIR-FIR colours give a comparison between the AGN torus versus host dust emission. 
We show  the observed 1.25$\mu$m - 65 $\mu$m, 1.25$\mu$m - 90 $\mu$m, 18 $\mu$m - 65 $\mu$m, 18 $\mu$m - 90 $\mu$m, 22 $\mu$m - 65 $\mu$m, 22 $\mu$m - 90 $\mu$m  
colours versus N$_{\rm{H}}$ in Fig. \ref{fig:fig9}. Since the median redshift of our sample is 0.02 we do not expect to have a significant redshift effect in the observed colours. 
As seen in Fig. \ref{fig:fig9} although there is a large scatter for individual colours, the median 1.25$\mu$m - 65 $\mu$m, 1.25$\mu$m - 90 $\mu$m, 18 $\mu$m - 65 $\mu$m, 18 $\mu$m - 90 $\mu$m, 22 $\mu$m - 65 $\mu$m, 
22 $\mu$m - 90 $\mu$m,  colours of the unobscured, obscured and CT AGN show an slightly increasing trend with N$_{\rm{H}}$.  
We note that we have checked other colour combinations and see a similar trend for $H$ - 65 $\mu$m, $H$ - 90 $\mu$m, $K_{s}$ - 65 $\mu$m, $K_{s}$ - 90 $\mu$m, 
12 $\mu$m- 65 $\mu$m and 12 $\mu$m- 90 $\mu$m. 
This positive trend indicates that the MIR-FIR colours gets slightly redder (or cooler) with increasing N$_{\rm{H}}$ values. 

\begin{figure*}
 \begin{center}$
 \begin{array}{cc}
 \includegraphics[]{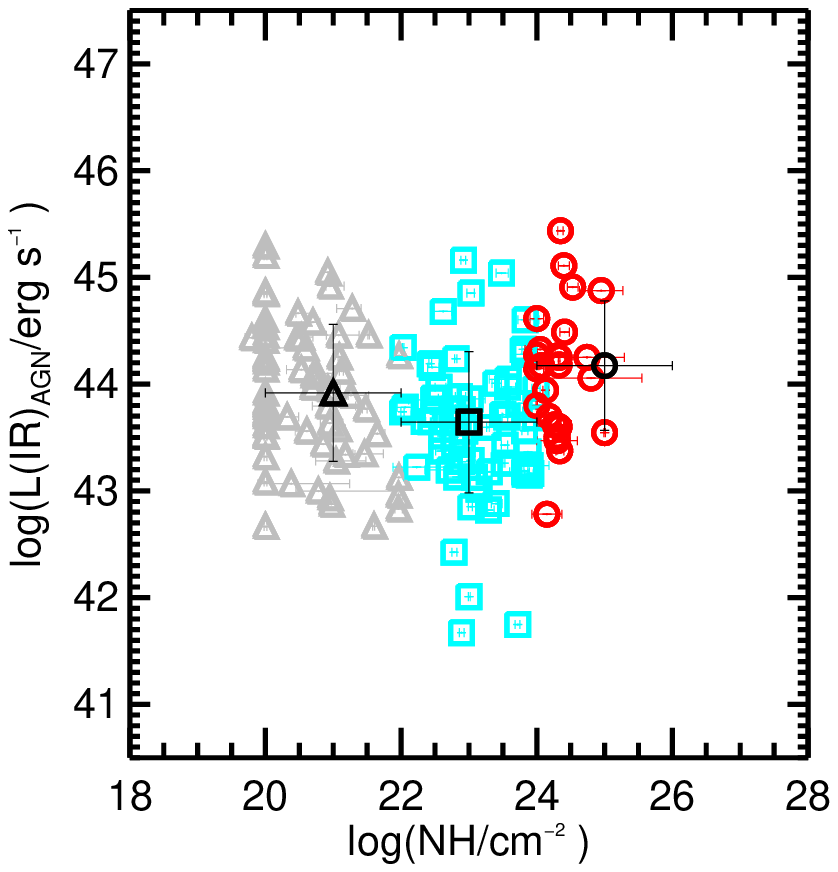}&
  \includegraphics[]{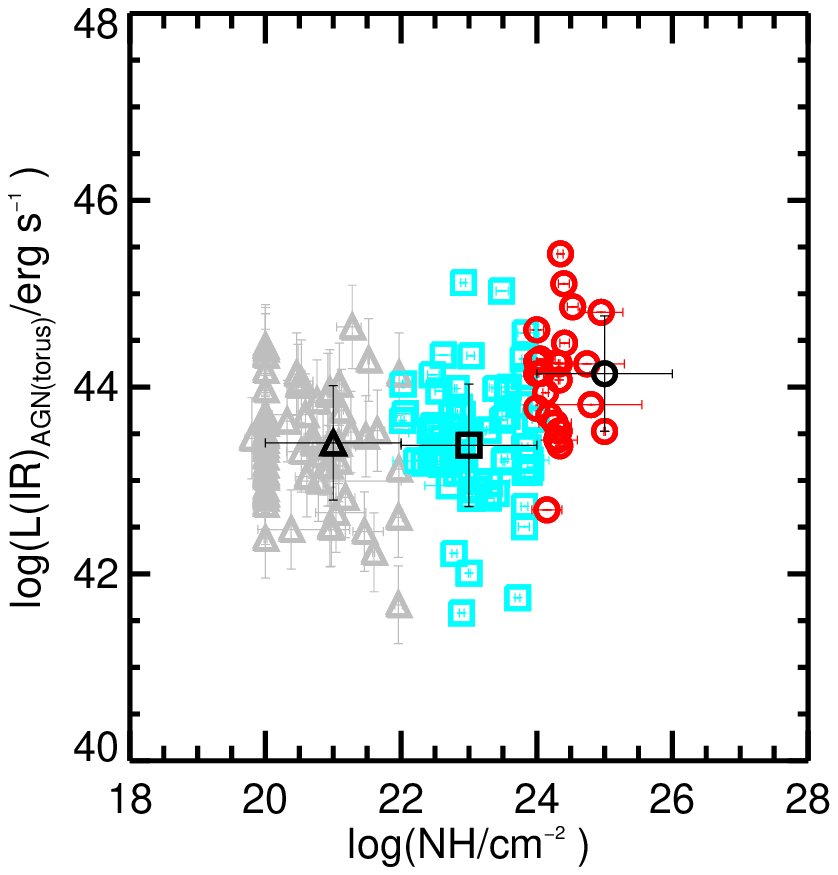}\\
 \end{array}$
 \end{center}
 \caption{ $L\rm{(IR)}_{\rm{AGN}}$ (left) and $L\rm{(IR)}_{\rm{AGN(torus)}}$ (right) versus N$_{\rm{H}}$. Compared to the unobscured and obscured AGN, median $L\rm{(IR)}_{\rm{AGN}}$ and $L\rm{(IR)}_{\rm{AGN(torus)}}$ 
 luminosities are higher for CT-AGN (black circle).} 
 \label{fig:fig10}
 \end{figure*}

 We have also investigated the $L\rm{(IR)}_{\rm{AGN}}$ dependence on the N$_{\rm{H}}$ value. As seen in the left panel of Fig. \ref{fig:fig10}, while unobscured, and 
 obscured AGN have a similar distribution CT-AGN have higher $L\rm{(IR)}_{\rm{AGN}}$  values. The right panel of  Fig. \ref{fig:fig10} shows that the infrared luminosity produced by the torus 
 ,$L\rm{(IR)}_{\rm{AGN(torus)}}$, is higher for  CT-AGN.

 \subsection{Obscured/Compton-thick AGN selection by IR colours} \label{S:selection}

\begin{figure}
\begin{center}$
\begin{array}{c}
\includegraphics[]{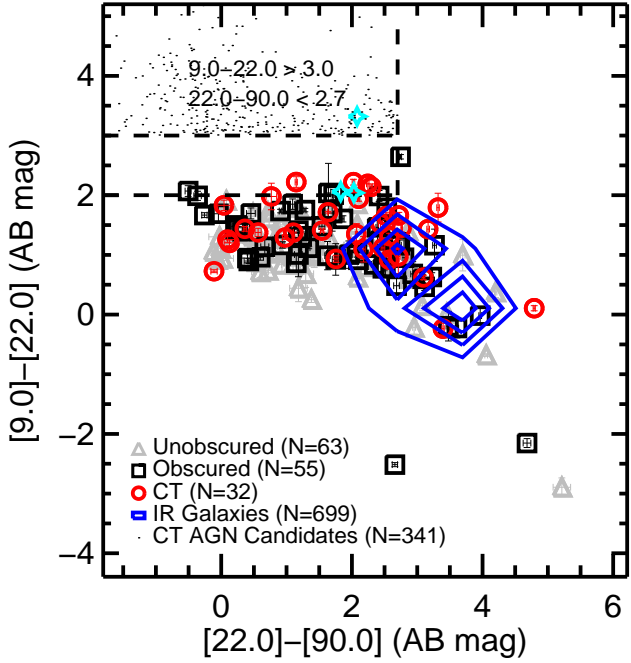}\\ 
\end{array}$
\end{center}
\caption{\textit{AKARI}$-$\textit{WISE} colours of X-ray selected AGNs with 9, 22, and 90 $\mu$m detections. 
The grey triangles are unobscured AGNs, the black squares are obscured AGNs and the red circles are Compton-thick AGNs. 
The blue contours represent 699 IR galaxies from \textit{AKARI} IR Galaxy catalogue of \citet{KilerciEserGoto2018} within the $z \leq$ 0.13 limit that have detections at 9, 22, and 90 $\mu$m bands. 
 Cyan diamonds represent the three CT AGN candidates selected from \textit{AKARI} IR galaxy catalogue of \citet{KilerciEserGoto2018} 
based on the  [$9\mu m - 22\mu m] > 2.0$ and [$22\mu m - 90\micron] < 2.7$ selection criteria. 
} 
\label{fig:fig11}
\end{figure}

We check many different colour-colour combinations including, optical, NIR, MIR and FIR, however we mostly find that CT AGNs 
distribute over a wide colour range similar to  obscured and unobscured AGNs. In one particular colour-colour diagram 
with 9 $\mu$m - 22 $\mu$m versus 22 $\mu$m - 90 $\mu$m colours (Fig. \ref{fig:fig11}) we find a distinct colour region which is mostly occupied by CT AGNs. 
This region is shown by the black dashed lines in Fig. \ref{fig:fig11}. In this figure the unobscured, obscured and CT AGNs are shown as the grey triangles, black squares and red circles, respectively. 
The blue contours are the 699 IR galaxies selected from the  \textit{AKARI} IR Galaxy catalogue of \citet{KilerciEserGoto2018} that have a similar redshift range of $z \leq$ 0.13. 
We define the [$9\mu m - 22\mu m] > 2.0$ and [$22\mu m - 90\micron] < 2.7$ as a new selection criteria for CT AGNs. 
The [$22\mu m - 90\mu m$]  colour separates AGN among IR galaxies, AGN have blue 22 $\mu$m - 90 $\mu$m colours. 
This is expected from the shallower slope of the $L_{\rm{dust}}$ and  $L\rm{(IR)}_{\rm{AGN}}$ relationship (section \ref{S:LxLir}.),  
considering the AGN IR emission traced by 22 $\mu$m and dust luminosity traced by 90 $\mu$m luminous AGN should give bluer  [$22\mu m - 90\mu m$] colours. 
\textit{AKARI}/IRC S9W band cover silicate absorption feature at 9.7$\mu$m. As shown by  \citet{Shi2006} the strength of 9.7$\mu$m absorption increases with higher N$_{\rm{H}}$. 
Therefore, the  [$9\mu m - 22\mu m$]  colour is expected to be related to deep silicate absorption feature seen in heavily obscured AGN. 
 22$\mu m$ is a measure of the MIR continuum. Once a source has a deep Silicate absorption, it will a have a fainter 9$\mu m$ and a larger difference in the [$22\mu m - 90\mu m$] colour.  
 In Fig. \ref{fig:fig11} we have four CT AGN (NGC6240, NGC7479, NGC6552, NGC5643) that is used to define [$9\mu m - 22\mu m] > 2.0$ colour range. \citet{Stierwalt2013} represent \textit{Spitzer} Infrared Spectrograph 
 \citep[IRS,][]{Houck2004} spectra covering 5-38$\mu$m of NGC6240. According to their silicate depth and MIR slope measurements, NGC6240 shows a strong silicate absorption with a steep MIR slope.
 \citet{Stone2016} measure the strength of the 9.7$\mu m$ silicate feature of NGC7479 from \textit{Spitzer} MIR spectra. According to their measurements, NGC7479 has a strong silicate absorption feature. 
 \citet{Shi2006} present \textit{Spitzer} IRS spectrum of NGC5643, measure a deep silicate absorption feature. 
  \textit{Spitzer} low-resolution IRS-LL spectrum of NGC6552 shows a rising MIR continuum between 15-35 $\mu m$  \citep{Jarrett2011}. 
  The \textit{Spitzer} spectrum of NGC6552 do not cover 9$\mu m$ range, therefore we do not  have a measure for its silicate absorption strength.
 The MIR spectral measurements of NGC6240, NGC7479 and NGC5643 show evidence for relatively faint 9$\mu m$ and relatively bright 22$\mu m$ giving higher  [$9\mu m - 22\mu m$] colours.  

We apply this criteria to the  \textit{AKARI} IR galaxy catalogue of \citet{KilerciEserGoto2018} and find 3 new CT AGNs candidates. These candidates are shown as cyan  diamonds in  Fig. \ref{fig:fig11}. 
As a first step, we justify the nature of the CT AGN candidates based on the available public data sets and previous studies. The properties of the CT AGN candidates selected from the \textit{AKARI} IR galaxy catalogue of \citet{KilerciEserGoto2018} is summarised in Table \ref{tab:table4}.

\begin{table*}
           \centering
           \small\addtolength{\tabcolsep}{-4pt}
           \caption{
           Properties of the CT AGN candidates. Columns: (1) Object name. (2) RA of the optical counterpart. (3) DEC of the optical counterpart.  
           (4) Spectroscopic redshift of the optical counterpart. (5) [9$\mu$m - 22$\mu$m] colour. (6) [22$\mu$m - 90$\mu$m] colour. (7)  Base 10 logarithm of the total infrared luminosity 
           between 8 -1000$\mu$m. (8) Base 10 logarithm of the intrinsic Hydrogen column density in units of cm$^{-2}$. (9) X-ray reference for the adopted $N_{\rm{H}}$ value in column (8). (10) Our comment about the  
           Compton-thick AGN candidate. 
	}
           \label{tab:table4}
            \begin{tabular}{lcccccclcc}
           \hline
           Name & RA           & Dec         & $z$        & [9$\mu$m - 22$\mu$m] & [22$\mu$m - 90$\mu$m]      & $\log (L_{IR}/L_{\sun})$          & $\log (N_{\rm{H}})$               & X-ray &Comment \\
                     & (J2000.0) &(J2000.0)  &                 &     [AB]                         &  [AB]                                     &                                            &  &  Ref.  & \\
                     & (deg)        & (deg)        &                 & (mag)                          & (mag)                                    &                                            &                                      &        & \\
            (1) & (2) & (3) & (4) & (5) & (6) & (7) & (8) & (9) & (10)  \\
           \hline
NGC\,1614    &   68.500 &  -8.579  & 0.016   &  2.04          & 2.03               & 11.60                &   21.58         &   \citet{PereiraSantaella2011} & uncertain AGN \\
NGC\,4418    & 186.727 &  -0.877  &0.007    &  3.32          & 2.08               & 10.94                &   $>$ 24       &      \citet{Maiolino2003}              & CT AGN \\
NGC\,7714    & 354.058 &  2.155   &0.009    &  2.06          & 1.83               & 10.72                &    21.34        &   \citet{Smith2005}   &  unobscured AGN \\ 
		\hline
	\end{tabular}
\end{table*}

NGC\,1614 is a well observed local LIRG. Its earlier X-ray observations with ASCA suggest the presence of an AGN based on the well fitted power-law in the 2-10\,keV band \citep{Risaliti2000}. 
Although the weak hard X-ray emission indicates a Compton-thick AGN at the first place, the multi-wavelength observations in sub-millimetre \citep{Xu2015}, radio \citep{HerreroIllana2017} and X-rays \citep{PereiraSantaella2011,HerreroIllana2014} do not support this. \citet{Xu2015} found that the amount of nuclear molecular gas and dust is much lower than that predicted for CT AGNs. 
According to \citet{PereiraSantaella2011,HerreroIllana2014} the hard and soft X-ray emission of this source can be explained by star-formation. 
As pointed out by \citep{PereiraSantaella2015}, in case of a CT AGN the predicted 14-195 keV flux would be above the  \textit{Swift}/BAT survey sensitivity limit. 
Therefore, the non-detection of this source in the \textit{Swift}/BAT survey \citep{Oh2018} supports that it is not an CT AGN.  
Based on the current multi-wavelength available data we consider it as an unlikely/uncertain CT AGN candidate. 
But we caution that NGC1614 has not been observed with NuSTAR yet, and its high energy nature is still a subject for investigation. 

NGC\,4418 is a LIRG with a known dust embedded nuclear CT source. The nature of this source has been subject to many observations at different wavelengths and the presence of a CT AGN is highly favoured. 
Infrared observations \citep{Roche1986,Spoon2001,Roche2015} of NGC\,4418 show a deep silicate absorption at 10\,$\mu$m and indicate a very heavily obscured, CT AGN. 
Submillimeter observations of NGC\,4418 at high spatial resolution  are also consistent with the presence of a CT AGN \citep{Sakamoto2013}. 
It's \textit{Chandra}/Advanced CCD Imaging Spectrometer (ACIS) observations show a flat hard X-ray spectrum and imply the presence of a CT AGN, but due to the limited photon statistics this identification is considered as tentative \citep{Maiolino2003}.
As pointed out by \citet{Roche2015}, the non-detection of NGC\,4418 by \textit{Swift}/BAT survey \citep{Baumgartner2013,Koss2013} is probably due to the extremely high levels of obscuration. 

NGC\,7714 is a well observed galaxy with a compact starburst nuclei \citep[e.g.,][]{Gonzalez-Delgado1995,Smith2005}. 
\citet{Smith2005} studied the X-ray emission from NGC\,7714 with \textit{Chandra}/ACIS. 
They report that the spectrum of the nuclear region can be fitted to a MEKAL function or an absorbed power-law function 
with a column density of N$_{\rm{H}}$ $= 2.2 \times10^{21}$ cm$^{-2}$. Therefore, NGC\,7714 possibly hosts an unobscured AGN. 

This investigation shows us that our new colour selection criteria is successful to find at least one confirmed CT AGN. Since 9 $\mu$m - 22 $\mu$m colour of NGC\,4418 is 3.32, we consider to take 
 [$9\mu m - 22\mu m] > 3.0$ as a more reliable region that would avoid unobscured AGN selection. When we apply   [$9\mu m - 22\mu m] > 3.0$ and [$22\mu m - 90\micron] < 2.7$ as a  criteria for a larger sample of combined \textit{AKARI} and \textit{WISE} 
sources without any selection criteria we find 341 candidates. 
These candidates are shown as the grey dots in Fig. \ref{fig:fig11}. 
We investigate the nature of these 341 candidates in astronomical databases such as HEASARC\footnote[7]{https://heasarc.gsfc.nasa.gov/cgi-bin/W3Browse/w3browse.pl}, 
NASA/IPAC Extragalactic Database\footnote[8]{http://nedwww.ipac.caltech.edu/.} (NED) and 
SIMBAD\footnote[9]{http://simbad.u-strasbg.fr/simbad/.} \citep{Wenger2000}. 
We find that most of these sources are Galactic IR sources such as young stellar objects (YSOs), asymptotic giant branch (AGB) stars, Planetary Nebulae (PNe)
and H\,{\sc ii}  and star-forming regions. Among these 341 sources only 3 (IRAS06190+1040, IRAS21144+5430, Mrk34) are classified as a galaxy in the searched databases. 
Mrk34 is found to be  a already known CT AGN in the literature \citep{Gandhi2014}.
However, while IRAS06190+1040 is classified as a galaxy in NED it is classified as a Galactic star cluster by  \citet{Froebrich2017}. 
IRAS21144+5430 is also classified as a galaxy in SDSS, but it is also classified as a H\,{\sc ii} region in SIMBAD. 
We also note that, IRAS06190+1040 and IRAS21144+5430 do not have archived X-ray observations for further investigation. 
We find 3 sources that are classified as IR sources with X-ray observations, these are  WISEA J162015.92-505621.8, 2MASS J16344841-4732432 and Allwise J181724.72-170628.1. 
X-ray observation of WISEA J162015.92-505621.8 show a weak X-ray source, but the quality of the available data does not allow for a reliable spectral analysis to measure N$_{\rm{H}}$. 
The available X-ray data of 2MASS J16344841-4732432 does not show a X-ray source. In the X-ray observation of Allwise J181724.72-170628.1 we identify a very weak X-ray source, 
the low S/N of the data does not allow to obtain a spectral fit. Based on the available X-ray data we can not confirm if these sources are CT AGN. However, all of these 3 sources are 
close to Galactic plane, and therefore they can also be Galactic IR sources.  

\section{Conclusions}\label{S:conc}

We have analysed the broad-band SEDs of 68 unobscured,  65 obscured and 25 CT AGN
from the UV to the far-IR by combining data from \textit{GALEX}, SDSS, 2MASS, \textit{WISE}, \textit{AKARI} and \textit{Herschel} surveys. 
We have constrained the parameters of the AGN component based on the AGN model \citet{Fritz2006}. 
We have investigated a possible IR colour selection criteria for CT AGN based on \textit{AKARI} and \textit{WISE} photometric magnitudes and found a new promising colour selection criteria. 
The results of this work can be summarised as follows: 

\begin{enumerate}

\item
The most important parameters to identify obscured/CT AGN from the SED analysis with CIGALE are the angle between equatorial axis and line of sight $\psi$, angular opening angle of the torus, $\theta$. 
SED models with $\psi = 0.001^{\circ}$ or $\psi = 10.0^{\circ}$, and  $\theta= 140$ , are very likely to indicate an obscured or a CT AGN. 

\item
The comparison of  average SEDs of the tree AGN populations show that the mid-IR SEDs are similar, the optical/UV region of the obscured/CT AGN is dominated 
by the host galaxy emission,  and in the far-infrared region CT AGN show a stronger emission.

\item
We identify 27 strong AGN luminosity dominated sources among the unobscured, obscured and CT AGN based on the $frac_\mathrm{AGN}$ obtained by the SED fitting. 
These represent AGN luminosity dominated UV to FIR SEDs. We find that while unobscured and obscured AGN have a similar $frac_\mathrm{AGN}$ distribution, 
CT AGN have higher $frac_\mathrm{AGN}$ values. 

\item
We find a moderately strong correlation luminosity correlations  between $L_{\rm{dust}}$-$L_{\rm{14 - 195}}$, $L\rm{(IR)}_{\rm{AGN}}$-$L_{\rm{14 - 195}}$ and
 $L\rm{(IR)}_{\rm{AGN}}$-$L_{\rm{14 - 195}}$. 
 We quantify the relationships between these separated IR luminosities and the ultra hard X-ray luminosity for the hard X-ray selected AGN from \textit{Swift}-BAT 105-month survey catalogue. 

\item
We compare average covering factor ($R= L\rm{(IR)}_{\rm{AGN}}$/$L\rm{(BOL)}_{\rm{AGN}}$) for unobscured, obscured and CT. We show that CT AGN have larger covering factors 
compared to the obscured and unobscured AGN. 

\item
We show that the median 1.25$\mu$m - 65 $\mu$m, 1.25$\mu$m - 90 $\mu$m, 18 $\mu$m - 65 $\mu$m, 18 $\mu$m - 90 $\mu$m, 22 $\mu$m - 65 $\mu$m, 22 $\mu$m - 90 $\mu$m
 colours of the unobscured, obscured and CT AGN have an increasing trend with N$_{\rm{H}}$. We show that MIR-FIR colours have a tendency to become redder (or cooler) with increasing N$_{\rm{H}}$. 
 
 \item
 We find that CT AGN have higher  $L\rm{(IR)}_{\rm{AGN}}$ and $L\rm{(IR)}_{\rm{AGN(torus)}}$ compared to obscured and unobscured AGN.
 
\item
We present a new CT AGN selection criteria as $9 - 22 > 3.0$ and $22 - 90 < 2.7$. 
As a result of this criteria we find two known CT AGN (NGC4418 and Mrk34) that are not included in  \textit{Swift}-BAT sample.  
Due to the limited number of sources detected in 9$\mu$m we could not find any new CT AGN.
We conclude that MIR photometric bands covering 9.7$\mu$m silicate absorption and MIR continuum can be used as a new tool to select the most heavily obscured CT AGN.

\end{enumerate}

\subsection*{Acknowledgements} 
 We thank the referee for many insightful comments. 
 EKE acknowledges a post-doctoral fellowship support from TUB\.{I}TAK-B\.{I}DEB through 2218 program. 
We thank Denis Burgarella for useful discussion about CIGALE. 
TG acknowledges the support by the Ministry of Science and Technology of Taiwan through grant NSC 108-2628-M-007-004-MY3. 
This research is based on observations with AKARI, a JAXA project with the participation of ESA. 
This publication makes use of data products from the Wide-field Infrared Survey Explorer, which is a joint project of the University of California, Los Angeles, and the Jet Propulsion Laboratory/California Institute of Technology, funded by the National Aeronautics and Space Administration. 
\textit{Herschel} is an ESA space observatory with science instruments
provided by European-led Principal Investigator consortia
and with important participation from NASA. 
This publication makes use of data products from the Two Micron All Sky Survey, which is a joint project of the University of Massachusetts and the Infrared Processing and Analysis Center/California Institute of Technology, funded by the National Aeronautics and Space Administration and the National Science Foundation. 


\bibliographystyle{mnras}
\bibliography{obsagn}


\appendix

\section{X-ray Spectral Analysis} \label{S:Appendix}

For \textit{XMM}-\textit{Newton} \citep{Jansen2001} we mostly used EPIC-pn camera data. 
Only for ESO450-16 we used both EPIC-pn and EPIC-MOS data.
We calibrated the data using SAS v15.0.0 \citep{Gabriel2004} and the calibration files as of 01/01/2019 using {\it epproc} package. 
We extracted source spectra using a circular region with a radius of about 32 arcsec. 
We extracted the background spectra from a nearby (located on the same CCD) uncontaminated larger circular region. 
Response and ancillary response files are generated using the {\it rmfgen} and {\it arfgen} tools. 

For one source we used \textit{Chandra}/ACIS \citep{Weisskopf2000,Garmire2003} data. 
\textit{Chandra} data were reduced using CIAO v.4.9 \citep{Fruscione2006} 
following standard procedures. We used {\it chandra$\_$repro} task to reprocess the data. 
The spectra were extracted using {\it specextract} tool from a circular aperture of 10 arcsec radius. 
Background spectra were extracted using a larger source free nearby circular region.

We performed X-ray spectral fitting using the \textsc{XSPEC} V. 12.10.1 software \citep{Arnaud1996}. 
We adopted Verner cross-section \citep{Verner1996} and Wilm abundance \citep{Wilms2000} in \textsc{TBABS} model to account for the Galactic absorption. 
Intrinsic absorption also modelled with  \textsc{TBABS}. 
We grouped all spectra to have at least 50 counts per channel (only for 2MASX J08384815+0407340 we used 40 counts) and used $\chi^{2}$ statistics. 
We fit the X-ray spectra in the 0.4 - 9.0 keV energy range, only in a few cases we had to narrow the used energy range due to the source brightness as 
shown in Fig. \ref{fig:fig12}.  
In order to measure the intrinsic N$_{\rm{H}}$ value the spectra are fitted using a single power-law modified by Galactic absorption plus an intrinsic absorption 
at the redshift of each source. 
During the spectral fitting process the power-law normalisation and the intrinsic column density were always left as free parameters. 
For 2MASXJ16510578-0129258, CGCG081-001 and MCG+00-58-028 
the slope of the power-law was left as a free parameter.  
As shown by \citet{Marchesi2018} fixing the photon index to the typical AGN value of $\Gamma \sim1.7$ in low-quality 2\,keV-10\,keV spectra gives 
better intrinsic N$_{\rm{H}}$ measurements. Therefore, for 2MASXJ08384815+0407340, IC-588, ESO439-G009, and NGC5940 we fixed photon index to 
$\Gamma=1.7$ or $\Gamma=1.8$. For ESO450-16 we limited the slope of the power-law between 1.0 and 2.0, for this source fixing $\Gamma$ to be 1.7 does not give a statistically acceptable fit. 
For these sources when necessary, we also added a Gaussian line to the power-law model  in order to obtain a better fit result. 
For IC-588, ESO439-G009 and NGC5940 we used the additional \textsc{mekal} model \citep{Mewe1985} to model the soft X-rays below 1\,keV. 
The obtained spectral parameters from the spectral fitting are listed in Table \ref{tab:tablex}. 
We show the best-fitting \textit{Chandra} and \textit{XMM}-\textit{Newton} X-ray spectra in Figure \ref{fig:fig12}.  
 As a result of our analysis we classify 2MASXJ08384815+0407340, IC-588, 2MASXJ16510578-0129258, MCG+00-58-028 as obscured AGN 
 and ESO439-G009, NGC5940, ESO450-16, CGCG081-001 as unobscured AGN.
 
 \begin{figure*}
 \begin{center}$
 \begin{array}{cc}
\includegraphics[]{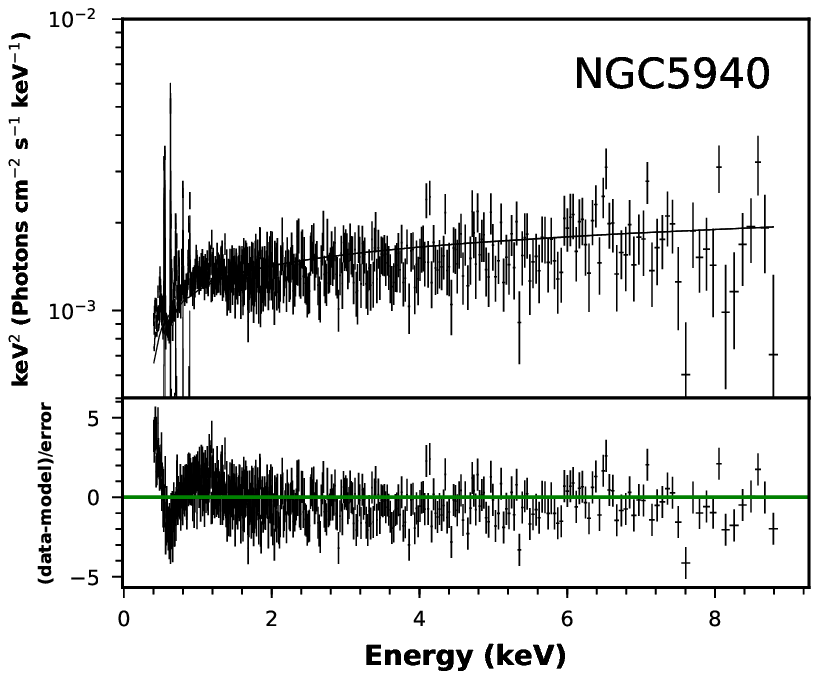} &
\includegraphics[]{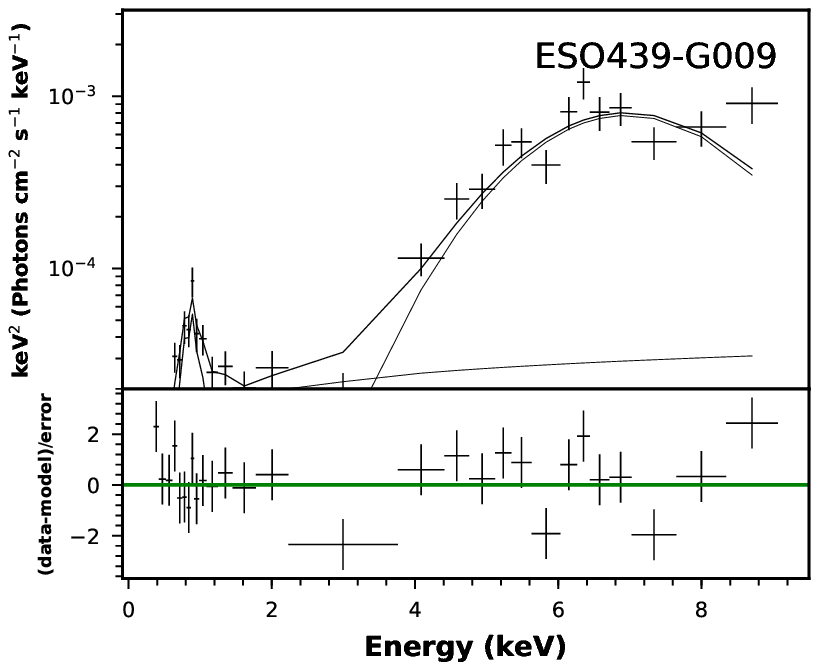} \\
\includegraphics[]{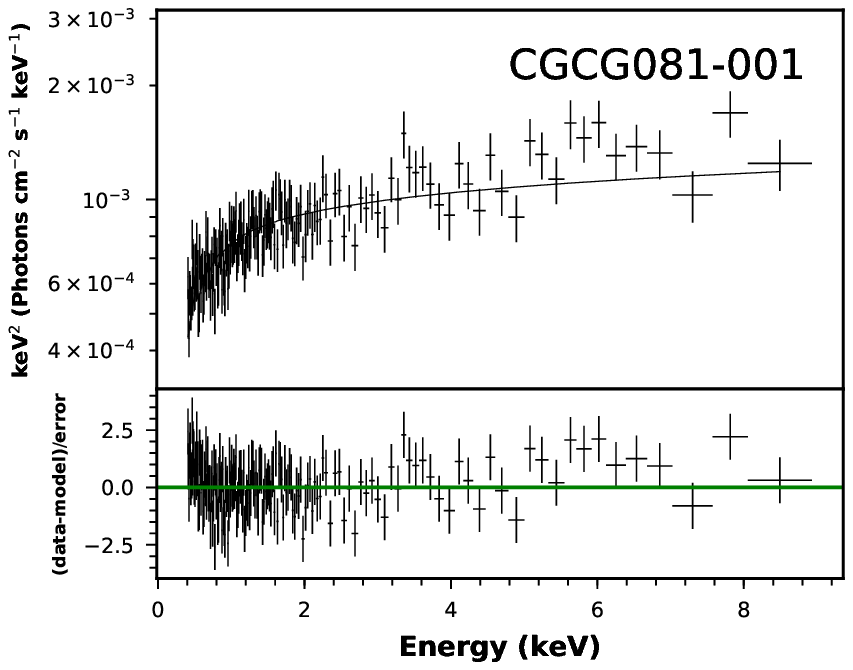} &
\includegraphics[]{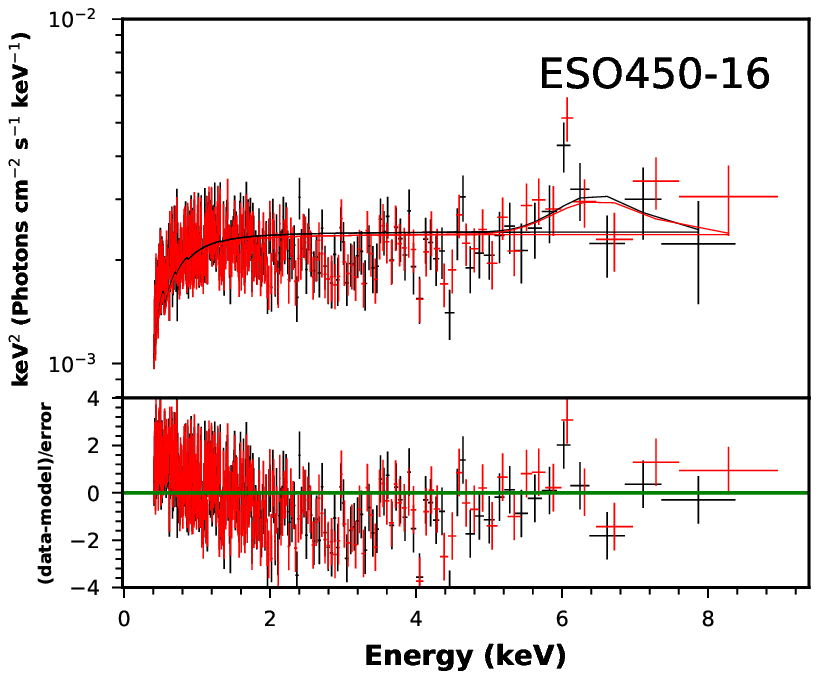} \\
 \includegraphics[]{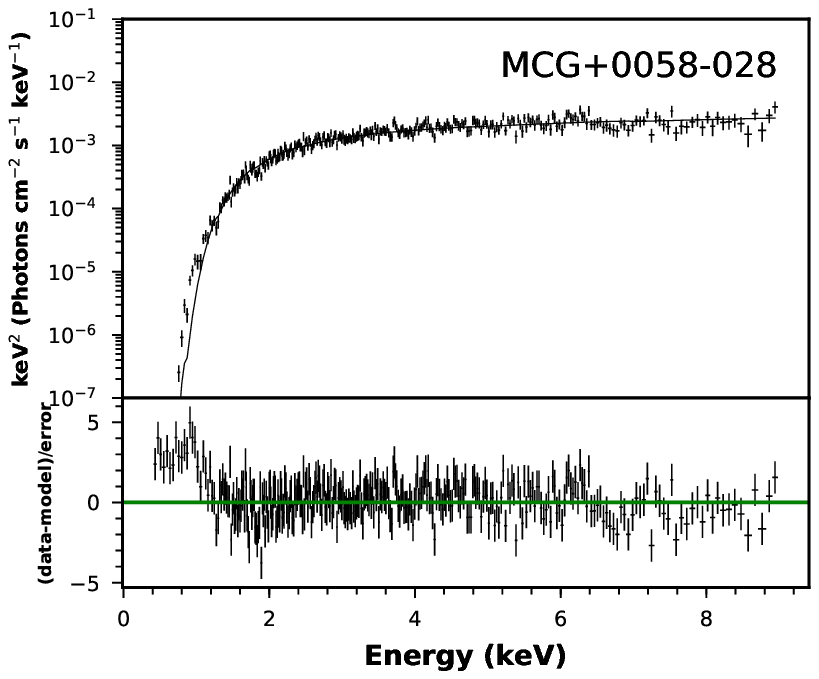} &
\includegraphics[]{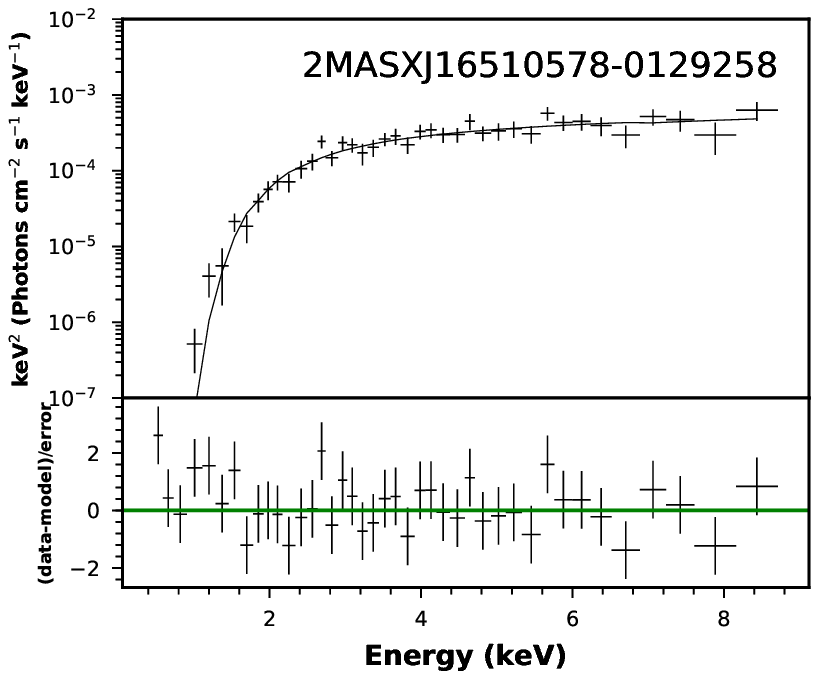} \\
 \end{array}$
 \end{center}
 \caption{ The best-fitting spectra of AGN observed with \textit{XMM}-\textit{Newton} and \textit{Chandra}.  
 For seven sources we use  \textit{XMM}-\textit{Newton} EPIC-pn data and only for ESO450-16 we used both EPIC-pn and EPIC-MOS (shown in red) data.}
 \label{fig:fig12}
 \end{figure*}
 
 \begin{figure*}
 \begin{center}$
 \begin{array}{cc}
\includegraphics[]{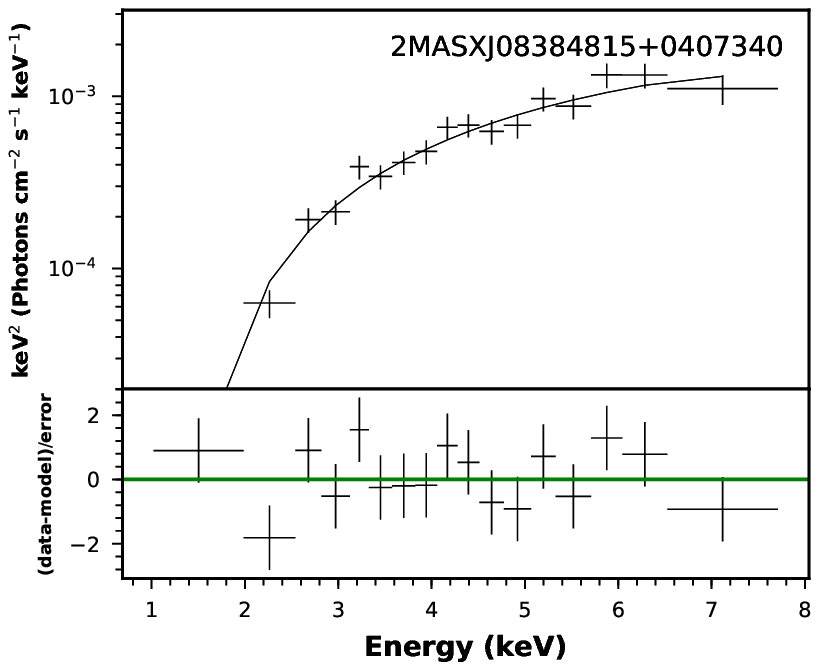} &
\includegraphics[]{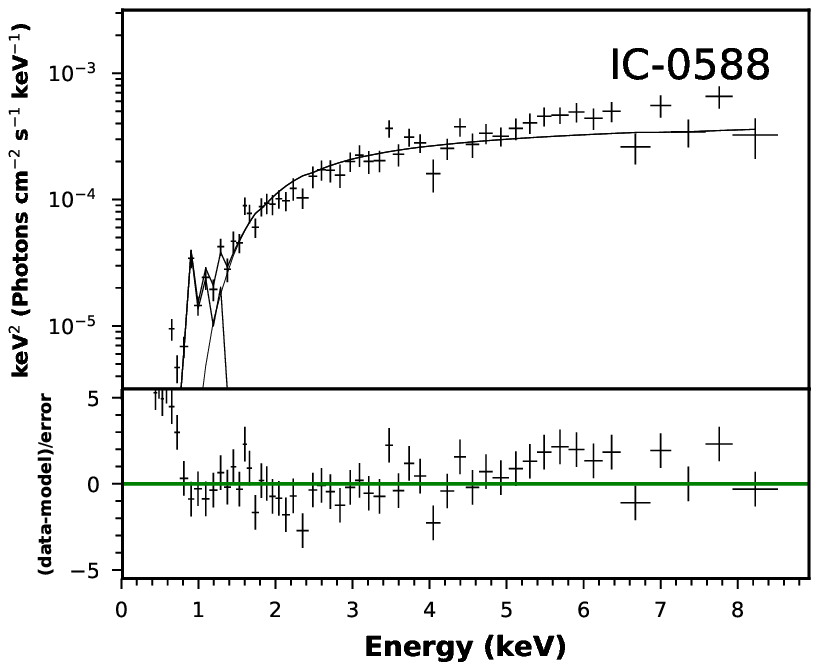} \\
\end{array}$
\end{center}
\contcaption{}
\end{figure*}
 
\begin{table*}
           \centering
           \caption{
           X-ray spectral analysis results. Columns: (1) Object name. (2) X-ray facility. (3) Galactic N$_{\rm{H}}$.   (4) Measured intrinsic N$_{\rm{H}}$. 
           (5) Temperature kT in keV of the \textsc{mekal} component. (6) Line energy of the simple gaussian profile in keV.  
           (7) Power-law photon index.  (8) Reduced chi-squared (degrees of freedom).
	}
           \label{tab:tablex}
            \begin{tabular}{lccccccc}
           \hline
           Source& Facility  & $N_{\rm{H}}\_{Gal}$      & $N_{\rm{H}}\_{int}$       & $kT$    &  Line energy   & $\Gamma$    & $\chi^{2}$ (d.o.f) \\
           name   &              &(10$^{20}$ cm$^{-2}$) &  (10$^{22}$ cm$^{-2}$)   &  (keV)  &    (keV)           &                       &                          \\
                       &              &                                     &                                        &             &                        &                      &                           \\
            (1)       & (2)        & (3) & (4) & (5) & (6) & (7) & (8)  \\
           \hline
2MASXJ08384815+0407340 &\textit{Chandra}                      & 2.68 &5.09$_{0.90}^{2.61}$                &                                  &0.20$_{2.45}^{0.90}$   & 1.7                               &1.20 (12)  \\
IC-588                                 & \textit{XMM}-\textit{Newton}     & 1.71 &2.88 $_{1.90}^{0.5}$                 &0.08$_{0.08}^{0.01}$ &                                    & 1.7                              & 4.34 (51)  \\
ES0439-G009                       &\textit{XMM}-\textit{Newton}    & 5.44 &$\leq 0.01$                                &0.68$_{0.05}^{0.07}$&6.40$_{0.15}^{0.17}$   & 1.7                              &1.88 (21)\\
NGC5940                               &\textit{XMM}-\textit{Newton}   & 3.95 &$\leq 0.01$                                &0.23$_{0.01}^{0.01}$ &                                     & 1.8                                & 1.70 (555)  \\
ESO450-16                            &\textit{XMM}-\textit{Newton}   & 7.46 &$\leq 0.01$                                &                                   &6.40$_{0.21}^{0.15}$  & 2.0$_{0.2}^{0.2}$       & 1.65 (558)    \\
CGCG081-001                       &\textit{XMM}-\textit{Newton}   & 4.43 &$\leq 0.01$                                &                                   &                                    & 1.83 $_{0.15}^{0.16}$  &1.14 (171) \\
2MASXJ16510578-0129258 &\textit{XMM}-\textit{Newton}	    & 7.80 &5.39$_{0.96}^{1.01}$                &                                   &                                     & 1.50 $_{0.22}^{0.23}$             &0.93 (38) \\
MCG+00-58-028                    &\textit{XMM}-\textit{Newton}    & 3.92 &3.29$_{0.12}^{0.12}$               &                                    &                                    & 1.59 $_{0.04}^{0.04}$            &1.63 (327) \\
		\hline
	\end{tabular}
\end{table*}



\bsp	
\label{lastpage}
\end{document}